\documentclass[foundations,article,submit,pdftex,moreauthors]{Definitions/mdpi} 
\firstpage{1} 
\pubvolume{1}
\issuenum{1}
\articlenumber{0}
\pubyear{2026}
\copyrightyear{2026}
\datereceived{ } 
\daterevised{ } 
\dateaccepted{ } 
\datepublished{ } 

\usepackage{bbm}

\newcommand{\pT}{\hat{+}}
\newcommand{\mT}{\hat{-}}
\newcommand{\muT}{\hat{\mu}}
\newcommand{\nuT}{\hat{\nu}}

\newcommand{\itP}[1]{\hat{#1}}

\usepackage{adjustbox}
\usepackage{amsmath}   
\usepackage{amssymb}   
\usepackage{booktabs}  
\usepackage{makecell}  
\usepackage{placeins}  
\Title{Interpolating conformal algebra in \texorpdfstring{$(3+1)$}{TEXT} dimensions between the instant form and the light-front form of relativistic dynamics}

\Author{Hariprashad Ravikumar $^{1}$\orcidA{} and Chueng-Ryong Ji $^{2}$\orcidB{}}

\AuthorNames{Hariprashad Ravikumar and Chueng-Ryong Ji}

\address{%
$^{1}$ \quad Department of Physics, New Mexico State University, Las Cruces, New Mexico 88003-8001, USA; hari1729@nmsu.edu\\
$^{2}$ \quad Department of Physics, North Carolina State University, Raleigh, North Carolina 27695-8202, USA; crji@ncsu.edu}

\corres{Correspondence: crji@ncsu.edu}

\abstract{We extend the interpolation of the Poincar\'e algebra between instant form dynamics (IFD) and light-front dynamics (LFD) to the conformal algebra in $(3+1)$ dimensions. Building upon our recent formulation of the interpolating conformal algebra in $(1+1)$ dimensions, we propose an interpolation method for the conformal group $SO(4,2)$. A central feature of this work is classifying the $15$ conformal generators into kinematic and dynamic categories as a function of the interpolation angle ($0 \leq \delta \leq \frac{\pi}{4}$). We show that among the five additional generators beyond the Poincar\'e group, the dilatation generator remains strictly kinematic across the entire interpolation region. We also find that in the exact light-front limit ($\delta = \frac{\pi}{4}$), one additional generator from the Special Conformal Transformations (SCT) becomes kinematic in $(1+1)$ but dynamical in $(3+1)$. This behavior shows the advantage of LFD in maximizing kinematic generators and minimizing dynamical complexity. To support this algebraic framework, we construct a $6 \times 6$ interpolating projective spacetime matrix representation. We detail the explicit transformations of the interpolating time under all $15$ generators, providing a systematic view of conformal symmetry across the relativistic dynamics.}


\begin{document}
\section{Introduction}
\label{sec:introduction}


Studying conformal symmetry is essential for discussing the fundamental properties of spacetime in relativistic quantum field theories~\cite{Wess1960, Kastrup1966, SalamMack1969, Gross1970} and developing characteristics of duality in the anti-de Sitter/conformal field theory (AdS/CFT) correspondence~\cite {Weinberg_2010, Brodsky_2015, Maldacena2016}. In the early 20th century, Cunningham  \cite{Cunningham1910} and Bateman  \cite{Bateman1910} demonstrated that the Maxwell equations are invariant under conformal transformations. Later, Dirac \cite{Dirac1936} proved that the massless version of his equation in relativistic quantum mechanics also exhibits invariance under conformal transformations. In the 1960s, Gross \cite{Gross1970} explored scale invariance as an asymptotic (high-energy) symmetry of scattering amplitudes, demonstrating that the scale invariance of Lagrangian field theories implies conformal invariance. A partial list of literature discussing the physical significance and potential applications of scale and conformal invariance can be found in Refs. \cite{Wess1960, Gürsey1956, Fulton1962, Kastrup1966, SalamMack1969, Wilson1969, Gross1970}. A historical review in Ref. \cite{Kastrup2008} provides further early references. Aspects of the conformal group in two-dimensional Minkowski spacetime, with representations of the algebra and of the group within a field-theoretic Schrödinger representation for bosons and fermions, were reviewed in Ref. \cite{Jackiw1990}.

In our previous work, we investigated the interpolating conformal algebra between instant-form dynamics (IFD) and light-front dynamics (LFD) in $(1+1)$ dimensions \cite{Ji-Hari-2026-PRD}. We constructed a $4\times4$ interpolating projective spacetime matrix representation to map the algebraic structures across all interpolation angles. We explored quantum mechanical operator representations by expressing the conformal generators in terms of the creation and annihilation operators of a simple harmonic oscillator, along with providing a $2\times2$ Pauli matrix representation for the $(1+0)$ and $(0+1)$ conformal groups. We demonstrated that out of the six generators in the $(1+1)$-dimensional conformal algebra, the number of kinematic generators is maximized from two in IFD to four in the LFD limit. This outcome signifies the utility of LFD by proving its capacity to save dynamical efforts in solving $(1+1)$-dimensional relativistic quantum field theories. Building on this foundation, we extend our analysis to $(3+1)$ dimensions in this work.

In general, the conformal transformation $x\longmapsto x'$ in $d$ dimensions can be defined as \cite{Francesco,Blumenhagen,Ji-Hari-2026-PRD},
\begin{align}
\frac{\partial x'^{\alpha}}{\partial x^{\mu}}\frac{\partial x'^{\beta}}{\partial x^{\nu}}g'_{\alpha\beta}=\Lambda(x)g_{\mu\nu},
\end{align}
meaning the metric is preserved up to a scale factor $\Lambda(x)$ under a conformal transformation, where $\mu,\nu\in\{0,\ldots,d-1\}$. The scale $\Lambda(x)=1$ corresponds to the Poincar\'e group consisting of translations, rotations, and Lorentz transformations. Consider an infinitesimal transformation, $x'^{\mu}=x^{\mu}+\epsilon^{\mu}(x)+\mathcal{O}(\epsilon^2)$. The metric then changes by $\delta g_{\mu\nu}=\partial_{\mu}\epsilon_{\nu}(x)+\partial_{\nu}\epsilon_{\mu}(x)$. The conformality condition requires
\begin{align}
\partial_{\mu}\epsilon_{\nu}(x)+\partial_{\nu}\epsilon_{\mu}(x)=F(x)\delta_{\mu\nu},\label{Killing}
\end{align}
where $F(x)=\Lambda(x)-1-\mathcal{O}(\epsilon^2)$. Equation~\eqref{Killing} is the conformal Killing equation. Contraction with $\delta^{\mu\nu}$ yields $F(x)=\frac{2}{d}\partial_{\mu}\epsilon^{\mu}$. There are four classes of solutions for $\epsilon_{\mu}(x)$:
\begin{align}
\label{conformal-generators}
P_{\mu}&=i\partial_{\mu} &&\text{(translations)},\nonumber\\
M_{\mu\nu}&=i\left(x_{\mu}\partial_{\nu}-x_{\nu}\partial_{\mu}\right) &&\text{(rotations and Lorentz transformations)},\nonumber\\
D&=ix_{\mu}\partial^{\mu} &&\text{(dilatation)},\nonumber\\
\mathfrak{K}_{\mu}&=i\left(2x_{\mu}x_{\nu}\partial^{\nu}-x^2\partial_{\mu}\right) &&\text{(special conformal transformations (SCT))}.
\end{align}
Their corresponding infinitesimal coordinate variations are $\epsilon^{\mu}(x)=a^{\mu}$, $\epsilon^{\mu}(x)=M^{\mu}_{~\nu}x^{\nu}$, $\epsilon^{\mu}(x)=\lambda x^{\mu}$, and $\epsilon^{\mu}(x)=2(b\cdot x)x^{\mu}-x^2b^{\mu}$, respectively. In finite form, the SCT is
\begin{align}
x'^{\mu}=\frac{x^{\mu}-b^{\mu}x^{2}}{1-2b\cdot x+b^{2}x^{2}},
\end{align}
which can be understood as an inversion of $x^\mu$, followed by a translation by $b^\mu$, and then another inversion \cite{Blumenhagen,Ji-Hari-2026-PRD}. These generators form the full conformal algebra, which contains the Poincar\'e group as a subgroup.
\begin{align}\label{conformalalgebra}
    \left[P_{{\mu}},P_{{\nu}}\right]&=0;~\left[\mathfrak{K}_{{\mu}},\mathfrak{K}_{{\nu}}\right]=0;\nonumber\\
 \left[D, P_{{\mu}}\right]&=iP_{{\mu}};~\left[D, \mathfrak{K}_{{\mu}}\right]=-i\mathfrak{K}_{{\mu}};\nonumber\\
 \left[P_{{\rho}},M_{{\mu}{\nu}}\right]&=i\left(g_{{\rho}{\mu}}P_{{\nu}}-g_{{\rho}{\nu}}P_{{\mu}}\right);\nonumber\\
 \left[\mathfrak{K}_{{\rho}},M_{{\mu}{\nu}}\right]&=i\left(g_{{\rho}{\mu}}\mathfrak{K}_{{\nu}}-g_{{\rho}{\nu}}\mathfrak{K}_{{\mu}}\right);\nonumber\\
 \left[M_{{\alpha}{\beta}},M_{{\rho}{\sigma}}\right]&=-i\left(g_{{\beta}{\sigma}}M_{{\alpha}{\rho}}-g_{{\beta}{\rho}}M_{{\alpha}{\sigma}}+g_{{\alpha}{\rho}}M_{{\beta}{\sigma}}-g_{{\alpha}{\sigma}}M_{{\beta}{\rho}}\right);\nonumber\\
 \left[\mathfrak{K}_{{\mu}},P_{{\nu}}\right]&=2i\left(g_{{\mu}{\nu}}D-M_{{\mu}{\nu}}\right);~\left[D, M_{{\mu}{\nu}}\right]=0.
\end{align}
where, $g_{\mu\nu}$ is the metric tensor.


In 1949, for the study of relativistic particle systems, Dirac \cite{Dirac1949} proposed three forms of relativistic dynamics: the instant form ($x^{0}=0$), the front form ($x^{+}=(x^{0}+x^{3})/\sqrt{2}$=0) and the point form ($x^{\mu}x_{\mu}=a^{2}>0, x^{0}>0$).
While the instant form dynamics (IFD) of quantum field theory is produced by quantization at equal time $t=x^{0}$, the light-front dynamics (LFD) is produced by quantization at equal light-front time $\tau \equiv (x^{0}+x^{3})/\sqrt{2}=x^{+}$ ($c=1$ unit is taken here).

One of the main reasons why LFD is advantageous over IFD can be attributed to the energy-momentum dispersion relation. For a particle with mass of $m$ and four-momentum of $k=(k^{0},k^{1},k^{2},k^{3})$, the irrational energy-momentum dispersion relation of the particle at equal-$t$ (IFD) is given by
\begin{align}
  k^{0}=\sqrt{\mathbf{k}^{2}+m^{2}}, \label{eqn:E-P_relation_IF}
\end{align}
where the energy $k^{0}$ is conjugate to $t$ and the three momentum vector $\mathbf{k}$ is given by $\mathbf{k}=(k^{1},k^{2},k^{3})$.
On the contrary, the corresponding rational energy-momentum relation at equal-$\tau$ (LFD) is given by
\begin{align}
  k^{-}=\dfrac{\mathbf{k}_{\perp}^{2}+m^{2}}{k^{+}}, \label{eqn:E-P_relation_LF}
\end{align}
where the light-front energy $k^{-}=(k^{0}-k^{3})/\sqrt{2}$ is conjugate to $\tau$, and the light-front momentum $k^{+}=(k^{0}+k^{3})/\sqrt{2}$ and $\mathbf{k}_{\perp}=(k^{1},k^{2})$ are the orthogonal momenta. This LFD energy-momentum relation is simpler and correlates the signs of $k^{-}$ and $k^{+}$.
When the system evolves in the future direction (positive $\tau$), a positive $k^{-}$ requires a positive $k^{+}$. This feature makes LFD distinct from other forms of relativistic Hamiltonian dynamics and prevents some processes from occurring. For example, spontaneous pair production from the vacuum is forbidden in LFD unless $k^{+}=0$ for both particles due to momentum conservation.

However, this sign correlation does not exist in IFD, and the vacuum structure appears much more complicated than in the LFD case due to the quantum fluctuations. 

\begin{table}[H]
  \caption{\label{tab:Kinematic_and_dynamic_generators_for_different_interoplation_angles}Kinematic and dynamic generators of the Poincaré group for different interpolation angles \cite{Ji2001, Ji2012, Ravikumar:2021xck}}
  \resizebox{\textwidth}{!}{%
    \begin{tabular}{lcc}
	& Kinematic & Dynamic \\
	\hline
	\rule{0pt}{3ex} $\delta=0$ & $\mathcal{K}^{\hat{1}}=-J^{2}, \mathcal{K}^{\hat{2}}=J^{1}, J^{3}, P^{1}, P^{2}, P^{3}$ & $\mathcal{D}^{\hat{1}}=-K^{1}, \mathcal{D}^{\hat{2}}=-K^{2}, -K_3, P^{0}$\\
	$0\leq\delta<\pi/4$ & $\mathcal{K}^{\hat{1}}, \mathcal{K}^{\hat{2}}, J^{3}, P^{1}, P^{2}, P_{\mT}$ & $\mathcal{D}^{\hat{1}}, \mathcal{D}^{\hat{2}}, -K_3, P_{\pT}$\\
	$\delta=\pi/4$ & $\mathcal{K}^{\hat{1}}=-E^{1}, \mathcal{K}^{\hat{2}}=-E^{2}, J^{3}, -K_3, P^{1}, P^{2}, P^{+}$ & $\mathcal{D}^{\hat{1}}=-F^{1}, \mathcal{D}^{\hat{2}}=-F^{2}, P^{-}$\\
      \end{tabular}%
  }
    \end{table}

Even the Poincar\'e algebra is drastically changed in the LFD compared to the IFD.
Among the ten Poincar\'e generators, we have the maximum number (seven) of kinematic (or interaction-independent) operators, which leave the state at $\tau=0$ unchanged. Indeed, the maximum number of kinematic generators allowed in relativistic dynamics is seven. The LFD is the only one that possesses this maximum number of kinematic generators.

LFD maximizes the capacity to
describe hadrons by saving dynamical effort in obtaining quantum chromodynamics (QCD) solutions that reflect the full Poincar\'e symmetries. Determining which generators are kinematic in each form of dynamics is useful, as having more kinematic generators saves dynamical effort when solving relativistic quantum field theories, such as quantum electrodynamics (QED) and QCD~\cite{ji2023relativistic}.

To interpolate the forms of relativistic quantum field theory between IFD and LFD, we take the following interpolating space-time coordinates  \cite{Ji2001, Hornbostel1992, Ji1996, Ji2012, Ji2015EM, Ji2015SP, Ji2018QED, Ji2021QCD, Ravikumar:2021xck} which is defined as a transformation from the ordinary space-time coordinates $x^{\muT}=\mathcal{R}^{\muT}_{\phantom{\mu}{\nu}}x^{\nu}$, i.e.,
\begin{align}\label{eqn:interpolation_angle_definition}
  \begin{pmatrix}
    x^{\hat{+}}\\
    x^{\hat{1}}\\
    x^{\hat{2}}\\
    x^{\hat{-}}
  \end{pmatrix}=
  \begin{pmatrix}
    \cos\delta & 0  & 0  & \sin\delta \\
    0          & 1  & 0  & 0 \\
    0          & 0  & 1  & 0 \\
    \sin\delta & 0  & 0  & -\cos\delta
  \end{pmatrix}
  \begin{pmatrix}
    x^{0}\\
    x^{1}\\
    x^{2}\\
    x^{3}
  \end{pmatrix},
\end{align}
where $\mathcal{R}^{\muT}_{\phantom{\mu}{\nu}}$ is given by the $4\times4$ involutory matrix with the determinant $-1$ with the interpolation angle allowed to run from $0^\circ$ (IFD) through $45^\circ$ (LFD), $0\le \delta \le \frac{\pi}{4}$. The interpolating coordinates $x^{\itP{\pm}}$ in the limit $\delta\rightarrow\pi/4$ become the light-front coordinates $x^{\pm}=(x^{0}\pm x^{3})/\sqrt{2}$ without  ``\textasciicircum''. Note that we interpolate from $-x^3$, so in the limit $\delta\rightarrow\pi/4$, the interpolating coordinates $x^{\itP{-}}$ is $-x^3$ (or $-z$) axis. Since the perpendicular components remain the same ($x^{\itP{j}}=x^{j},x_{\itP{j}}=x_{j}, j=1,2$), we will omit the ``\textasciicircum''  notation unless necessary from now on for the perpendicular indices $j=1,2$ in a four-vector.

 In this interpolating basis, the metric becomes
\begin{align}\label{eqn:g_munu_interpolation}
  g^{\hat{\mu}\hat{\nu}}
  = g_{\hat{\mu}\hat{\nu}}
  =
  \begin{pmatrix}
    \mathbb{C} & 0  & 0  & \mathbb{S} \\
    0 & -1 & 0  & 0 \\
    0 & 0  & -1 & 0 \\
    \mathbb{S} & 0  & 0  & -\mathbb{C}
  \end{pmatrix},
\end{align}
where $\mathbb{S}=\sin2\delta$ and $\mathbb{C}=\cos2\delta$. The lower index variables $x_{\hat{+}}$ and $x_{\hat{-}}$ are related to the upper index variables as $x_{\hat{+}}=g_{\hat{+}\hat{+}}x^{\hat{+}}+g_{\hat{+}\hat{-}}x^{\hat{-}}=\mathbb{C}x^{\hat{+}}+\mathbb{S}x^{\hat{-}}$ and $x_{\hat{-}}=g_{\hat{-}\hat{+}}x^{\hat{+}}+g_{\hat{-}\hat{-}}x^{\hat{-}}=-\mathbb{C}x^{\hat{-}}+\mathbb{S}x^{\hat{+}}$.
The details of the relationship between the
interpolating variables and the usual space-time variables
can be found in our previous works  \cite{Ji2001, Ravikumar:2021xck, Hornbostel1992, Ji1996, Ji2012, Ji2015EM, Ji2015SP, Ji2018QED, Ji2021QCD, Ji-Hari-2026-PRD}.

To interpolate the Lorentzian tensor for the Poincar\'e algebra, we start from the tensor in IFD given by 
\begin{align}\label{eqn:J_mu_nu_IF}
  M_{\mu\nu}=\begin{pmatrix}
     0&-K^1&-K^2&K_3\\
     K^1&0&J^3&-J^2\\
     K^2&-J^3&0&J^1\\
     -K_3&J^2&-J^1&0 
     \end{pmatrix} ,
\end{align}
and apply the unitary-type transformation with the involutory matrix given by Eq.(\ref{eqn:interpolation_angle_definition}), namely
\begin{align}\label{eqn:Poincare_Matrix_Interpolation_superscripts}
  M_{\muT\nuT}
  &=\mathcal{R}_{\hat{\mu}}^{{\alpha}}M_{\alpha\beta}\mathcal{R}_{\hat{\nu}}^{{\beta}}=\begin{pmatrix}
    0 & {\mathcal{D}}^{\itP{1}} & {\mathcal{D}}^{\itP{2}} & -K_3\\
    -{\mathcal{D}}^{\itP{1}} & 0 & {J}^{3} & -{\mathcal{K}}^{\itP{1}}\\
    -{\mathcal{D}}^{\itP{2}} & -{J}^{3} & 0 & -{\mathcal{K}}^{\itP{2}}\\
    K_3 & {\mathcal{K}}^{\itP{1}} & {\mathcal{K}}^{\itP{2}} & 0
  \end{pmatrix},
\end{align}
and
\begin{align}
  M^{\muT\nuT}
  =
  g^{\muT\itP{\alpha}}M_{\itP{\alpha}\itP{\beta}}g^{\itP{\beta}\nuT}
  =
  \begin{pmatrix}
    0 & {E}^{\itP{1}} & {E}^{\itP{2}} & K_3\\
    -{E}^{\itP{1}} & 0 & {J}^{3} & -{F}^{\itP{1}}\\
    -{E}^{\itP{2}} & -{J}^{3} & 0 & -{F}^{\itP{2}}\\
    -K_3 & {F}^{\itP{1}} & {F}^{\itP{2}} & 0
  \end{pmatrix} ,
\end{align}
where
\begin{align}\label{eqn:E_F_D_K_Definition_Interpolation_Angle}
  &E^{\itP{1}}=J^{2}\sin\delta+K^{1}\cos\delta,
  &&\mathcal{K}^{\itP{1}}=-K^{1}\sin\delta-J^{2}\cos\delta, \nonumber\\
  &E^{\itP{2}}=K^{2}\cos\delta-J^{1}\sin\delta,
  &&\mathcal{K}^{\itP{2}}=J^{1}\cos\delta-K^{2}\sin\delta, \nonumber\\
  &F^{\itP{1}}=K^{1}\sin\delta-J^{2}\cos\delta,
  &&\mathcal{D}^{\itP{1}}=-K^{1}\cos\delta+J^{2}\sin\delta, \nonumber\\
  &F^{\itP{2}}=K^{2}\sin\delta+J^{1}\cos\delta,
  &&\mathcal{D}^{\itP{2}}=-J^{1}\sin\delta-K^{2}\cos\delta.
\end{align}
In the limit $\delta=\pi/4$, the interpolating $E^{\itP{j}}$ and $F^{\itP{j}}$ will coincide with the usual $E^{j}$ and $F^{j}$ of LFD.
Note here that the ``\textasciicircum'' notation is reinstated for $j=1, 2$ to emphasize the angle $\delta$ dependence and that the position of the indices on $K, J, E, F, \mathcal{D}, \mathcal{K}$ won't matter, as they are not four-vectors: i.e., $F_{\itP{1}}=F^{\itP{1}}$, etc.

The generalized Poincar\'e algebra for any interpolation angle can be found in \cite{Ji2001, Ravikumar:2021xck}. 
Among the ten Poincar\'e generators, the six generators ($\mathcal{K}^{\itP{1}}, \mathcal{K}^{\itP{2}}, J^{3}, P_{1}, P_{2}, P_{\mT}$) are always kinematic in the sense that the $x^{\pT}=0$ plane remains intact under the transformations generated by them.
As discussed in \cite{Ji2001, Ji2012, Ravikumar:2021xck}, the operator $-K_3=M_{\pT\mT}$ is dynamical in the region where $0\leq\delta<\pi/4$ but becomes kinematic in the light-front limit ($\delta=\pi/4$). Therefore, the instant defined by $x^+=0$ becomes invariant under longitudinal boosts as we move to the light front.
The set of kinematic and dynamic generators depending on the interpolation angle is summarized in Table.~\ref{tab:Kinematic_and_dynamic_generators_for_different_interoplation_angles}.
Since the kinematic transformations do not alter $x^{\pT}$, the individual time-ordered amplitude must be invariant under the kinematic transformations \cite{Ji2018QED}.

In the following section, Section~\ref{sec:conformal-ifd}, we extend this interpolation method to the conformal group, which carries the Poincar\'e group as a subgroup. We then discuss the interpolation method for the conformal group $SO(4,2)$. To characterize the individual properties of the conformal generators, the $6\times6$ interpolating projective spacetime matrix representation is constructed in Section~\ref{sec:projectivespace}, where we systematically classify the kinematic and dynamic generators across the interpolation angles. Summary and conclusion follow in Section~\ref{summary-conclusion}.
In Appendix~\ref{app:explicit_matrices}, 
we present the correspondence of explicit Dirac matrices with the generators of the conformal algebra.
In Appendix \ref{app:interpolating-time-derivations}, we detail the derivation of the interpolating time transformations for all conformal generators given in 
Table~\ref{tablexprime}.


\section{From Poincar\'e Algebra to Conformal Algebra}
\label{sec:conformal-ifd}
As summarized in Eq.(\ref{conformal-generators}), the conformal algebra include five more generators ($D, \mathfrak{K}_\mu$) beyond the ten Poincar\'e generators represented by the $4\times 4$ Poincar\'e matrix $M_{\mu,\nu}$ and the translation generators $P_\mu$. 
We may thus rather naturally extend the $4\times4$ Poincar\'e matrix given by Eq.(\ref{eqn:J_mu_nu_IF}) in IFD to the 
following antisymmetric $6\times6$ array \cite{Francesco,Blumenhagen,Ji-Hari-2026-PRD} including the column and row vectors of the translation generators $P_\mu$:
\begin{align}\label{Jab}
  J_{ab}&=\begin{pmatrix}
      0&-D&\frac{-\mathfrak{K}_{\mu}}{\sqrt{2}}\\
      D&0&\frac{P_{\mu}}{\sqrt{2}}\\
      \frac{\mathfrak{K}_{\mu}}{\sqrt{2}}&\frac{-P_{\mu}}{\sqrt{2}}&M_{\mu\nu}
  \end{pmatrix} \nonumber \\
  &=
  \begin{pmatrix}
  0&-D&\frac{-\mathfrak{K}_0}{\sqrt{2}}&\frac{-\mathfrak{K}_1}{\sqrt{2}}&\frac{-\mathfrak{K}_2}{\sqrt{2}}&\frac{-\mathfrak{K}_3}{\sqrt{2}}\\
  D&0&\frac{P_0}{\sqrt{2}}&\frac{P_1}{\sqrt{2}}&\frac{P_2}{\sqrt{2}}&\frac{P_3}{\sqrt{2}}\\
    \frac{\mathfrak{K}_0}{\sqrt{2}}&\frac{-P_0}{\sqrt{2}}&0 & -K^{1} & -K^{2} & K_3\\
    \frac{\mathfrak{K}_1}{\sqrt{2}}&\frac{-P_1}{\sqrt{2}}&K^{1} & 0 & J^{3} & -J^{2}\\
    \frac{\mathfrak{K}_2}{\sqrt{2}}&\frac{-P_2}{\sqrt{2}}&K^{2} & -J^{3} & 0 & J^{1}\\
    \frac{\mathfrak{K}_3}{\sqrt{2}}&\frac{-P_3}{\sqrt{2}}&-K_3 & J^{2} & -J^{1} & 0
  \end{pmatrix}_{6\times6} ,
\end{align}
where $J_{ab}=-J_{ba}$ and $a,b\in\{-2,-1,0,1,2,3\}$. We note here that the two extra dimensions denoted as $-2$ and $-1$ are added to the typical (3+1) dimensional space-time coordinates 
$\{0,1,2,3\}$ enlarging the typical four-dimensional space-time base manifold of the relativistic dynamics into the six-dimensional conformal manifold. Then, these fifteen conformal algebraic generators 
obey the $SO(4,2)$ commutation relations:
  \begin{align}\label{algebrasimp}
      \left[J_{{a}{b}},J_{{c}{d}}\right]=-i\left(g_{{b}{d}}J_{{a}{c}}-g_{{b}{c}}J_{{a}{d}}+g_{{a}{c}}J_{{b}{d}}-g_{{a}{d}}J_{{b}{c}}\right) 
  \end{align}
where, 
  \begin{align}\label{metric}
      g_{ab}=\begin{pmatrix}
  0&-1&0&0&0&0\\
  -1&0&0&0&0&0\\
  0&0&1&0&0&0\\
  0&0&0&-1&0&0\\
  0&0&0&0&-1&0\\
  0&0&0&0&0&-1\\
  \end{pmatrix}_{6\times6} .
  \end{align}
The algebra in Eq.~\eqref{algebrasimp}, with the metric in Eq.~\eqref{metric}, is the IFD realization of the conformal algebra. In particular,
$J_{-2,-1}=-D$, $J_{-2,\mu}=-\mathfrak{K}_{\mu}/\sqrt{2}$, and
$J_{-1,\mu}=P_{\mu}/\sqrt{2}$. Its explicit component commutators are summarized in Table~\ref{tableIFD}. \\

\begingroup
\fontsize{7}{8}\selectfont
\setlength{\tabcolsep}{1pt}
\renewcommand{\arraystretch}{1.25}
\noindent
\begin{minipage}[t]{\textwidth}
\centering
\adjustbox{max width=\textwidth, max totalheight=0.55\textheight}{%
\begin{tabular}{ |c||c|c|c|c|c|c|c|c|c|c|c|c|c|c|c| }
 \hline
 \rule{0pt}{16pt} & $\mathfrak{K}_{{0}}$ & $P_{{0}}$ & $D$ & $\mathfrak{K}_{{3}}$ & $P_{{3}}$ & $K_{{3}}$ & $P_{{1}}$ & $P_{{2}}$ & $\mathfrak{K}_{{1}}$ & $\mathfrak{K}_{{2}}$  & ${K}^{{1}}$ & ${K}^{{2}}$  & ${J}^{{2}}$ & ${J}^{{1}}$ & $J^{{3}}$ \\
 \hline
  \hline
 \rule{0pt}{16pt} $\mathfrak{K}_{{0}}$ &0&$-2iD$&$-i\mathfrak{K}_{{0}}$&0&$-2iK_{{3}}$&${i\mathfrak{K}_{{3}}}$&$2i{K}^{{1}}$&$2i{K}^{{2}}$&0&0&$-i\mathfrak{K}_{{1}}$&$-i\mathfrak{K}_{{2}}$&0&0&0\\
 \hline 
 \rule{0pt}{16pt} $P_{{0}}$ &$2iD$&0&$iP_{{0}}$&${-2iK_{{3}}}$&0&${iP_{{3}}}$&0&0&$2i{K}^{{1}}$&$2i{K}^{{2}}$&$-iP_{{1}}$&$-iP_{{2}}$&0&0&0\\
 \hline 
 \rule{0pt}{16pt} $D$ &$i\mathfrak{K}_{{0}}$&$-iP_{{0}}$&0&$i\mathfrak{K}_{{3}}$&$-iP_{{3}}$&0&$-iP_{{1}}$&$-iP_{{2}}$&$i\mathfrak{K}_{{1}}$&$i\mathfrak{K}_{{2}}$&0&0&0&0&0\\
 \hline 
 \rule{0pt}{16pt} $\mathfrak{K}_{{3}}$ &0&${2iK_{{3}}}$&${-i\mathfrak{K}_{{3}}}$&0&$2iD$&${i\mathfrak{K}_{{0}}}$&${-2i{J}^{{2}}}$&${2i{J}^{{1}}}$&0&0&0&0&${-i\mathfrak{K}_{{1}}}$&${i\mathfrak{K}_{{2}}}$&0\\
 \hline 
 \rule{0pt}{16pt} $P_{{3}}$ &$2iK_{{3}}$&0&$iP_{{3}}$&$-2iD$&0&${iP_{{0}}}$&0&0&${-2i{J}^{{2}}}$&${2i{J}^{{1}}}$&0&0&${-iP_{{1}}}$&${iP_{{2}}}$&0\\
 \hline 
 \rule{0pt}{16pt} $K_{{3}}$ &${-i\mathfrak{K}_{{3}}}$&${-iP_{{3}}}$&0&${-i\mathfrak{K}_{{0}}}$&${-iP_{{0}}}$&0&0&0&0&0&$i{J}^{{2}}$&$-i{J}^{{1}}$&${i{K}^{{1}}}$&$-{i{K}^{{2}}}$&0\\
 \hline 
 \rule{0pt}{16pt} $P_{{1}}$ &$-2i{K}^{{1}}$&0&$iP_{{1}}$&${2i{J}^{{2}}}$&0&0&0&0&$-2iD$&$-2iJ^{{3}}$&$-iP_{{0}}$&0&${iP_{{3}}}$&0&$-iP_{{2}}$\\
 \hline 
 \rule{0pt}{16pt} $P_{{2}}$ &$-2i{K}^{{2}}$&0&$iP_{{2}}$&${-2i{J}^{{1}}}$&0&0&0&0&$2iJ^{{3}}$&$-2iD$&0&$-iP_{{0}}$&0&${-iP_{{3}}}$&$iP_{{1}}$\\
 \hline 
 \rule{0pt}{16pt} $\mathfrak{K}_{{1}}$ &0&$-2i{K}^{{1}}$&$-i\mathfrak{K}_{{1}}$&0&${2i{J}^{{2}}}$&0&$2iD$&$-2iJ^{{3}}$&0&0&$-i\mathfrak{K}_{{0}}$&0&${i\mathfrak{K}_{{3}}}$&0&$-i\mathfrak{K}_{{2}}$\\
 \hline 
 \rule{0pt}{16pt} $\mathfrak{K}_{{2}}$ &0&$-2i{K}^{{2}}$&$-i\mathfrak{K}_{{2}}$&0&${-2i{J}^{{1}}}$&0&$2iJ^{{3}}$&$2iD$&0&0&0&$-i\mathfrak{K}_{{0}}$&0&${-i\mathfrak{K}_{{3}}}$&$i\mathfrak{K}_{{1}}$\\
 \hline 
 \rule{0pt}{16pt} ${K}^{{1}}$ &$i\mathfrak{K}_{{1}}$&$iP_{{1}}$&0&0&0&$-i{J}^{{2}}$&$iP_{{0}}$&0&$i\mathfrak{K}_{{0}}$&0&0&$-iJ^{{3}}$&${-iK_{{3}}}$&0&$-i{K}^{{2}}$\\
 \hline 
 \rule{0pt}{16pt} ${K}^{{2}}$ &$i\mathfrak{K}_{{2}}$&$iP_{{2}}$&0&0&0&$i{J}^{{1}}$&0&$iP_{{0}}$&0&$i\mathfrak{K}_{{0}}$&$iJ^{{3}}$&0&0&${iK_{{3}}}$&$i{K}^{{1}}$\\
 \hline 
 \rule{0pt}{16pt} ${J}^{{2}}$ &0&0&0&${i\mathfrak{K}_{{1}}}$&${iP_{{1}}}$&$-i{K}^{{1}}$&$-iP_{{3}}$&0&${-i\mathfrak{K}_{{3}}}$&0&$iK_{{3}}$&0&0&$-iJ^{{3}}$&$i{J}^{{1}}$\\
 \hline 
 \rule{0pt}{16pt} ${J}^{{1}}$ &0&0&0&${-i\mathfrak{K}_{{2}}}$&${-iP_{{2}}}$&$i{K}^{{2}}$&0&$iP_{{3}}$&0&${i\mathfrak{K}_{{3}}}$&0&$-iK_{{3}}$&$iJ^{{3}}$&0&$-i{J}^{{2}}$\\
 \hline 
 \rule{0pt}{16pt} $J^{{3}}$ &0&0&0&0&0&0&$iP_{{2}}$&$-iP_{{1}}$&$i\mathfrak{K}_{{2}}$&$-i\mathfrak{K}_{{1}}$&$i{K}^{{2}}$&$-i{K}^{{1}}$&$-i{J}^{{1}}$&$i{J}^{{2}}$&0\\
 \hline 
\end{tabular}} 
\captionof{table}{Conformal algebra $(3+1)$ in the IFD. }
\label{tableIFD}
\end{minipage}
\endgroup




Now, we extend the interpolating Poincar\'e algebra discussed in the previous section, Section \ref{sec:introduction}, to the interpolating conformal algebra between IFD and LFD.  
To find the $6\times6$ projective spacetime representations in the interpolating form, it is convenient to extend the $4\times4$ involutory matrix $\mathcal{R}^{\muT}_{\phantom{\mu}{\nu}}$ introduced in Eq.(\ref{eqn:interpolation_angle_definition}) to the  
 $6\times6$ involutary transformation matrix $(\mathcal{R}_{\hat{a}}^{\phantom{b}{b}})_{6\times6}$ given by
\begin{align}
    (\mathcal{R}_{\hat{a}}^{\phantom{b}{b}})_{6\times6}=(\mathcal{R}_{\hat{a}}^{\phantom{b}{b}})^T_{6\times6}=\begin{pmatrix}
    1&0&0&0&0&0\\
    0&1&0&0&0&0\\
    0&0&\cos{\delta}&0&0&\sin{\delta}\\
    0&0&0&1&0&0\\
    0&0&0&0&1&0\\
    0&0&\sin{\delta}&0&0&-\cos{\delta}
    \end{pmatrix}
\end{align}
and apply it to Eq.\eqref{Jab}. Then, the interpolation form of the antisymmetric tensor of conformal generators, $J_{\hat{a}\hat{b}}$, can be found from the unitary-type transformation given by  $J_{\hat{a}\hat{b}}=\mathcal{R}_{\hat{a}}^{~{c}}J_{cd}\mathcal{R}_{~\hat{b}}^{{d}}$. The explicit computation leads to $J_{\hat{a}\hat{b}}$ as given by
\begin{align}
    J_{\hat{a}\hat{b}}&=\begin{pmatrix}
    0&-D&-\frac{\mathfrak{K}^{\hat{+}}}{\sqrt{2}}&-\frac{\mathfrak{K}_1}{\sqrt{2}}&-\frac{\mathfrak{K}_2}{\sqrt{2}}&-\frac{\mathfrak{K}^{\hat{-}}}{\sqrt{2}}\\
    D&0&\frac{P_{\hat{+}}}{\sqrt{2}}&\frac{P_{1}}{\sqrt{2}}&\frac{P_{2}}{\sqrt{2}}&\frac{P_{\hat{-}}}{\sqrt{2}}\\
    \frac{\mathfrak{K}^{\hat{+}}}{\sqrt{2}}&-\frac{P_{\hat{+}}}{\sqrt{2}}&0 & {\mathcal{D}}^{\itP{1}} & {\mathcal{D}}^{\itP{2}} & -K_3\\
    \frac{\mathfrak{K}_1}{\sqrt{2}}&-\frac{P_{1}}{\sqrt{2}}&-{\mathcal{D}}^{\itP{1}} & 0 & {J}^{3} & -{\mathcal{K}}^{\itP{1}}\\
    \frac{\mathfrak{K}_2}{\sqrt{2}}&-\frac{P_{2}}{\sqrt{2}}&-{\mathcal{D}}^{\itP{2}} & -{J}^{3} & 0 & -{\mathcal{K}}^{\itP{2}}\\
    \frac{\mathfrak{K}^{\hat{-}}}{\sqrt{2}}&-\frac{P_{\hat{-}}}{\sqrt{2}}&K_3 & {\mathcal{K}}^{\itP{1}} & {\mathcal{K}}^{\itP{2}} & 0
  \end{pmatrix}_{(6\times6)}
  \label{Jhat+hat-} ,
\end{align}
where $J_{\hat{a}\hat{b}}=-J_{\hat{b}\hat{a}}$ and $\hat{a},\hat{b}\in\{-2,-1,\hat{+},1,2,\hat{-}\}$.  
The simplified interpolating conformal algebra is then given by
  \begin{align}
      \left[J_{{\hat{a}}{\hat{b}}},J_{{\hat{c}}{\hat{d}}}\right]=-i\left(g_{{\hat{b}}{\hat{d}}}J_{{\hat{a}}{\hat{c}}}-g_{{\hat{b}}{\hat{c}}}J_{{\hat{a}}{\hat{d}}}+g_{{\hat{a}}{\hat{c}}}J_{{\hat{b}}{\hat{d}}}-g_{{\hat{a}}{\hat{d}}}J_{{\hat{b}}{\hat{c}}}\right),
      \label{simplesrint}
  \end{align}
where
\begin{align}
    g_{\hat{a}\hat{b}}&=\begin{pmatrix}
  0&-1&0&0&0&0\\
  -1&0&0&0&0&0\\
  0&0&\mathbb{C}&0&0&\mathbb{S}\\
  0&0&0&-1&0&0\\
  0&0&0&0&-1&0\\
  0&0&\mathbb{S}&0&0&-\mathbb{C}\\
  \end{pmatrix}_{6\times6}.\label{metricghat}
\end{align}
While the dilatation generator $D = J_{-1-2}$ and the longitudinal boost generator $K_3 = J_{{\hat -}{\hat +}}$ are separately given as particular components of $J_{{\hat a}{\hat b}}$ in Eq.(\ref{Jhat+hat-}),
they can be naturally mixed to each other as shown in our previous work~\cite{Ji-Hari-2026-PRD}, namely
\begin{align}
\label{mixingDandK3}
    \begin{pmatrix}
        D_{\hat{+}}\\
        D_{\hat{-}}
    \end{pmatrix}=&\begin{pmatrix}
        \cos\delta&\sin\delta\\
        \sin\delta&-\cos\delta
    \end{pmatrix}\begin{pmatrix}
        D\\
        K_3
    \end{pmatrix}.
\end{align}
Such mixing of $D$ and $K_3$ follows the same pattern of the interpolation between $P_0$ and $P_3$ as well as between $\mathfrak{K}_0$ and $\mathfrak{K}_3$ as the $4\times 4$ matrix representation of $J_{ab}$ ($a,b \in \{-2,-1,0,3\}$) in (1+1)d conformal algebra can be split into two separate $3\times 3$ matrix representations of $J_{pq}^{(0)} (p,q \in \{1,2,3\})$ and  $J_{rs}^{(3)} (r,s \in \{1,2,3\})$ in which the superscripts $(0)$ and $(3)$ signify the involvement of time and space transformations respectively in (1+1)d. The more details of this formulation in (1+1)d conformal algebra should be referred to Ref.\cite{Ji-Hari-2026-PRD}. It may be understandable from the physics point of view that the dilatation generator $D$ and the longitudinal boost generator $K_3$ are the same sort of conformal generators that can be mixed each other as the dilatation of space and time can be realizable just like the length contraction and the time dilation attained in the longitudinal boost. 

In Table~\ref{tab:generator_limits}, we summarize the correspondence of each and every 15 interpolating generators of the conformal algebra with the ones in IFD and LFD, taking the limit $\delta \to 0$ and $\delta \to \pi/4$, respectively. 
Note that the interpolation in projective spacetime naturally produces superscript special conformal generators  ($\mathfrak{K}^{\hat{\pm}}$), which are related to the subscript special conformal generators as $\mathfrak{K}^{\hat{\pm}}=g^{\hat{\pm}\hat{\pm}}\mathfrak{K}_{\hat{\pm}}+g^{\hat{\pm}\hat{\mp}}\mathfrak{K}_{\hat{\mp}}$ with the metric explicitly defined in Eq.\eqref{eqn:g_munu_interpolation}, i.e.
\begin{align}
    \begin{pmatrix}
        \mathfrak{K}^{\hat{+}}\\
        \mathfrak{K}^{\hat{-}}
    \end{pmatrix}=&\begin{pmatrix}
        \mathbb{C}&\mathbb{S}\\
        \mathbb{S}&-\mathbb{C}
    \end{pmatrix}\begin{pmatrix}
        \mathfrak{K}_{\hat{+}}\\
        \mathfrak{K}_{\hat{-}}
    \end{pmatrix}.
\end{align}
Then, the algebra in Eq.\eqref{simplesrint}, with the above interpolating $6\times6$ metric, reproduces $_{15}C_{2} = 105$ explicit commutation relations in the interpolating conformal algebra between IFD and LFD. These 105 explicit commutation relations are 
presented in Table~\ref{tableinterpolation} with the abbreviated notations 
$s=\sin{\delta}$ and $c=\cos{\delta}$ as well as $\mathbb{S}=\sin2\delta$,  $\mathbb{C}=\cos2\delta$. In the IFD limit $c\to 1~\& \, s\to 0$ as well as 
$\mathbb{C}\rightarrow1~\&~\mathbb{S}\rightarrow0$, one can see that Table~\ref{tableinterpolation} reproduced Table~\ref{tableIFD}.  
Likewise, in the LFD limit $c\to 0~\& \, s\to 1$ as well as 
$\mathbb{C}\rightarrow0~\&~\mathbb{S}\rightarrow1$, Table~\ref{tableinterpolation} yields Table~\ref{tableLFD} which lists the explicit commutation relations of the conformal algebra in LFD. 

We note that the full $15 \times 15$ interpolating algebra presented in Table~\ref{tableinterpolation} reduces to the 
$6 \times 6$ interpolating algebra presented in our previous work~\cite{Ji-Hari-2026-PRD} of the $(1+1)$-dimensional interpolating conformal algebra. In Table III and Table IV of Ref.~\cite{Ji-Hari-2026-PRD}, one can find the respective (1+1)-dimensional reduced conformal algebra of IFD and LFD corresponding to Table~\ref{tableIFD}
and Table~\ref{tableLFD} of the present work.
In the same vein of the correspondence presented in this work, Table III and Table IV of Ref.~\cite{Ji-Hari-2026-PRD} can be respectively found in the IFD and LFD limit 
from the interpolating algebra presented in 
Table V of Ref.~\cite{Ji-Hari-2026-PRD}. 

In the next section, Section \ref{sec:projectivespace}, we extend the projective spacetime representations of the 6 generators presented in our previous 
work~\cite{Ji-Hari-2026-PRD} of the $(1+1)$-dimensional interpolating conformal
algebra to the 15 generators of the $(3+1)$-dimensional interpolating conformal
algebra. We present explicitly the projective spacetime representations of all 15 interpolating generators between IFD and LFD.

\begin{table}[htpb!]
\centering
\renewcommand{\arraystretch}{2} 
\caption{The 15 conformal generators in the interpolation form alongside their Instant-Form Dynamics (IFD) and Light-Front Dynamics (LFD) limits.}
\label{tab:generator_limits}
\begin{tabular}{@{}ccc@{}}
\toprule
\makecell{\textbf{Interpolating} \\ \textbf{generators}} & 
\makecell{\textbf{IFD limit} \\ ($\delta \to 0$)} & 
\makecell{\textbf{LFD limit} \\ ($\delta \to \frac{\pi}{4}$)} \\
\midrule

$\begin{aligned}
P_{\hat{+}} &= P_0\cos\delta + P_3\sin\delta \\
P_{\hat{-}} &= P_0\sin\delta - P_3\cos\delta
\end{aligned}$ &
$\begin{aligned}
P_{\hat{+}} &\to P_0 \\
P_{\hat{-}} &\to -P_3
\end{aligned}$ &
$\begin{aligned}
P_{\hat{+}} &\to P_{+} = \frac{P_0 + P_3}{\sqrt{2}} \\
P_{\hat{-}} &\to P_{-} = \frac{P_0 - P_3}{\sqrt{2}}
\end{aligned}$ \\

\midrule

$\begin{aligned}
\mathfrak{K}_{\hat{+}} &= \mathfrak{K}_0\cos\delta - \mathfrak{K}_3\sin\delta \\
\mathfrak{K}_{\hat{-}} &= \mathfrak{K}_0\sin\delta + \mathfrak{K}_3\cos\delta
\end{aligned}$ &
$\begin{aligned}
\mathfrak{K}_{\hat{+}} &\to \mathfrak{K}_0 \\
\mathfrak{K}_{\hat{-}} &\to \mathfrak{K}_3
\end{aligned}$ &
$\begin{aligned}
\mathfrak{K}_{\hat{+}} &\to \mathfrak{K}_{+} = \frac{\mathfrak{K}_0 - \mathfrak{K}_3}{\sqrt{2}} \\
\mathfrak{K}_{\hat{-}} &\to \mathfrak{K}_{-} = \frac{\mathfrak{K}_0 + \mathfrak{K}_3}{\sqrt{2}}
\end{aligned}$ \\

\midrule

$\begin{aligned}
D_{\hat{+}} &= D\cos\delta + K_3\sin\delta \\
D_{\hat{-}} &= D\sin\delta - K_3\cos\delta
\end{aligned}$ &
$\begin{aligned}
D_{\hat{+}} &\to D \\
D_{\hat{-}} &\to -K_3
\end{aligned}$ &
$\begin{aligned}
D_{\hat{+}} &\to D_{+} = \frac{D + K_3}{\sqrt{2}} \\
D_{\hat{-}} &\to D_{-} = \frac{D - K_3}{\sqrt{2}}
\end{aligned}$ \\

\midrule

$\begin{aligned}
\mathcal{K}^{\hat{1}} &= -K^1\sin\delta - J^2\cos\delta \\
\mathcal{K}^{\hat{2}} &= J^1\cos\delta - K^2\sin\delta
\end{aligned}$ &
$\begin{aligned}
\mathcal{K}^{\hat{1}} &\to -J^2 \\
\mathcal{K}^{\hat{2}} &\to J^1
\end{aligned}$ &
$\begin{aligned}
\mathcal{K}^{\hat{1}} &\to -E^1 = \frac{-K^1 - J^2}{\sqrt{2}} \\
\mathcal{K}^{\hat{2}} &\to -E^2 = \frac{J^1 - K^2}{\sqrt{2}}
\end{aligned}$ \\

\midrule

$\begin{aligned}
\mathcal{D}^{\hat{1}} &= -K^1\cos\delta + J^2\sin\delta \\
\mathcal{D}^{\hat{2}} &= -J^1\sin\delta - K^2\cos\delta
\end{aligned}$ &
$\begin{aligned}
\mathcal{D}^{\hat{1}} &\to -K^1 \\
\mathcal{D}^{\hat{2}} &\to -K^2
\end{aligned}$ &
$\begin{aligned}
\mathcal{D}^{\hat{1}} &\to -F^1 = \frac{-K^1 + J^2}{\sqrt{2}} \\
\mathcal{D}^{\hat{2}} &\to -F^2 = \frac{-J^1 - K^2}{\sqrt{2}}
\end{aligned}$ \\

\midrule

$J^3 = J^3$ & $J^3 \to J^3$ & $J^3 \to J^3$ \\
$P_{1,2} = P_{1,2}$ & $P_{1,2} \to P_{1,2}$ & $P_{1,2} \to P_{1,2}$ \\
$\mathfrak{K}_{1,2} = \mathfrak{K}_{1,2}$ & $\mathfrak{K}_{1,2} \to \mathfrak{K}_{1,2}$ & $\mathfrak{K}_{1,2} \to \mathfrak{K}_{1,2}$ \\

\bottomrule
\end{tabular}
\end{table}

\startlandscape
\begingroup
\fontsize{8}{9}\selectfont
\setlength{\tabcolsep}{1pt}
\renewcommand{\arraystretch}{2.5}
\noindent
\begin{minipage}[t]{\textwidth}
\centering
\captionof{table}{Full $15 \times 15$ conformal algebra in the interpolation form}
\label{tab:full_interpolation}
\resizebox{\textwidth}{!}{%
\begin{tabular}{ |c||c|c|c|c|c|c|c|c|c|c|c|c|c|c|c| }
\hline
\rule{0pt}{16pt}  & $\mathfrak{K}_{\hat{+}}$ & $P_{\hat{+}}$ & $D_{\hat{+}}$ & $\mathfrak{K}_{\hat{-}}$ & $P_{\hat{-}}$ & $D_{\hat{-}}$ & $P_{1}$ & $P_{2}$ & $\mathfrak{K}_{1}$ & $\mathfrak{K}_{2}$ & $\mathcal{D}^{\hat{1}}$ & $\mathcal{D}^{\hat{2}}$ & $\mathcal{K}^{\hat{1}}$ & $\mathcal{K}^{\hat{2}}$ & $J^{3}$ \\
\hline
\hline
\rule{0pt}{16pt} $\mathfrak{K}_{\hat{+}}$ & $0$ & $-2i[(c+\mathbb{S}s)d_{\hat{+}}+(s-\mathbb{S}c)d_{\hat{-}}]$ & $-i[(c+\mathbb{S}s)\mathfrak{K}_{\hat{+}}+(s-\mathbb{S}c)\mathfrak{K}_{\hat{-}}]$ & $0$ & $2i\mathbb{C}[sD_{\hat{+}}-cD_{\hat{-}}]$ & $i\mathbb{C}[s\mathfrak{K}_{\hat{+}}-c\mathfrak{K}_{\hat{-}}]$ & $-2i[\mathbb{S}\mathcal{K}^{\hat{1}} + \mathbb{C}\mathcal{D}^{\hat{1}}]$ & $-2i[\mathbb{S}\mathcal{K}^{\hat{2}} + \mathbb{C}\mathcal{D}^{\hat{2}}]$ & $0$ & $0$ & $i\mathfrak{K}_{1}$ & $i\mathfrak{K}_{2}$ & $0$ & $0$ & $0$ \\
\hline
\rule{0pt}{16pt} $P_{\hat{+}}$ & $2i[(c+\mathbb{S}s)d_{\hat{+}}+(s-\mathbb{S}c)d_{\hat{-}}]$ & $0$ & $i[(c+\mathbb{S}s)P_{\hat{+}}+(s-\mathbb{S}c)P_{\hat{-}}]$ & $-2i\mathbb{C}[sD_{\hat{+}}-cD_{\hat{-}}]$ & $0$ & $-i\mathbb{C}[sP_{\hat{+}}-cP_{\hat{-}}]$ & $0$ & $0$ & $-2i\mathcal{D}^{\hat{1}}$ & $-2i\mathcal{D}^{\hat{2}}$ & $i\mathbb{C} P_{1}$ & $i\mathbb{C} P_{2}$ & $i\mathbb{S} P_{1}$ & $i\mathbb{S} P_{2}$ & $0$ \\
\hline
\rule{0pt}{16pt} $D_{\hat{+}}$ & $i[(c+\mathbb{S}s)\mathfrak{K}_{\hat{+}}+(s-\mathbb{S}c)\mathfrak{K}_{\hat{-}}]$ & $-i[(c+\mathbb{S}s)P_{\hat{+}}+(s-\mathbb{S}c)P_{\hat{-}}]$ & $0$ & $-i\mathbb{C}[s\mathfrak{K}_{\hat{+}}-c\mathfrak{K}_{\hat{-}}]$ & $i\mathbb{C}[sP_{\hat{+}}-cP_{\hat{-}}]$ & $0$ & $-ic P_{1}$ & $-ic P_{2}$ & $ic\mathfrak{K}_{1}$ & $ic\mathfrak{K}_{2}$ & $is[\mathbb{C}\mathcal{K}^{\hat{1}} - \mathbb{S}\mathcal{D}^{\hat{1}}]$ & $is[\mathbb{C}\mathcal{K}^{\hat{2}} - \mathbb{S}\mathcal{D}^{\hat{2}}]$ & $is[\mathbb{S}\mathcal{K}^{\hat{1}} + \mathbb{C}\mathcal{D}^{\hat{1}}]$ & $is[\mathbb{S}\mathcal{K}^{\hat{2}} + \mathbb{C}\mathcal{D}^{\hat{2}}]$ & $0$ \\
\hline
\rule{0pt}{16pt} $\mathfrak{K}_{\hat{-}}$ & $0$ & $2i\mathbb{C}[sD_{\hat{+}}-cD_{\hat{-}}]$ & $i\mathbb{C}[s\mathfrak{K}_{\hat{+}}-c\mathfrak{K}_{\hat{-}}]$ & $0$ & $-2i[(c-\mathbb{S}s)d_{\hat{+}}+(s+\mathbb{S}c)d_{\hat{-}}]$ & $-i[(c-\mathbb{S}s)\mathfrak{K}_{\hat{+}}+(s+\mathbb{S}c)\mathfrak{K}_{\hat{-}}]$ & $2i[\mathbb{C}\mathcal{K}^{\hat{1}} - \mathbb{S}\mathcal{D}^{\hat{1}}]$ & $2i[\mathbb{C}\mathcal{K}^{\hat{2}} - \mathbb{S}\mathcal{D}^{\hat{2}}]$ & $0$ & $0$ & $0$ & $0$ & $i\mathfrak{K}_{1}$ & $i\mathfrak{K}_{2}$ & $0$ \\
\hline
\rule{0pt}{16pt} $P_{\hat{-}}$ & $-2i\mathbb{C}[sD_{\hat{+}}-cD_{\hat{-}}]$ & $0$ & $-i\mathbb{C}[sP_{\hat{+}}-cP_{\hat{-}}]$ & $2i[(c-\mathbb{S}s)d_{\hat{+}}+(s+\mathbb{S}c)d_{\hat{-}}]$ & $0$ & $i[(c-\mathbb{S}s)P_{\hat{+}}+(s+\mathbb{S}c)P_{\hat{-}}]$ & $0$ & $0$ & $-2i\mathcal{K}^{\hat{1}}$ & $-2i\mathcal{K}^{\hat{2}}$ & $i\mathbb{S} P_{1}$ & $i\mathbb{S} P_{2}$ & $-i\mathbb{C} P_{1}$ & $-i\mathbb{C} P_{2}$ & $0$ \\
\hline
\rule{0pt}{16pt} $D_{\hat{-}}$ & $-i\mathbb{C}[s\mathfrak{K}_{\hat{+}}-c\mathfrak{K}_{\hat{-}}]$ & $i\mathbb{C}[sP_{\hat{+}}-cP_{\hat{-}}]$ & $0$ & $i[(c-\mathbb{S}s)\mathfrak{K}_{\hat{+}}+(s+\mathbb{S}c)\mathfrak{K}_{\hat{-}}]$ & $-i[(c-\mathbb{S}s)P_{\hat{+}}+(s+\mathbb{S}c)P_{\hat{-}}]$ & $0$ & $-is P_{1}$ & $-is P_{2}$ & $is\mathfrak{K}_{1}$ & $is\mathfrak{K}_{2}$ & $-ic[\mathbb{C}\mathcal{K}^{\hat{1}} - \mathbb{S}\mathcal{D}^{\hat{1}}]$ & $-ic[\mathbb{C}\mathcal{K}^{\hat{2}} - \mathbb{S}\mathcal{D}^{\hat{2}}]$ & $-ic[\mathbb{S}\mathcal{K}^{\hat{1}} + \mathbb{C}\mathcal{D}^{\hat{1}}]$ & $-ic[\mathbb{S}\mathcal{K}^{\hat{2}} + \mathbb{C}\mathcal{D}^{\hat{2}}]$ & $0$ \\
\hline
\rule{0pt}{16pt} $P_{1}$ & $2i[\mathbb{S}\mathcal{K}^{\hat{1}} + \mathbb{C}\mathcal{D}^{\hat{1}}]$ & $0$ & $ic P_{1}$ & $-2i[\mathbb{C}\mathcal{K}^{\hat{1}} - \mathbb{S}\mathcal{D}^{\hat{1}}]$ & $0$ & $is P_{1}$ & $0$ & $0$ & $-2i[cD_{\hat{+}} + sD_{\hat{-}}]$ & $-2iJ^{3}$ & $i P_{\hat{+}}$ & $0$ & $i P_{\hat{-}}$ & $0$ & $-i P_{2}$ \\
\hline
\rule{0pt}{16pt} $P_{2}$ & $2i[\mathbb{S}\mathcal{K}^{\hat{2}} + \mathbb{C}\mathcal{D}^{\hat{2}}]$ & $0$ & $ic P_{2}$ & $-2i[\mathbb{C}\mathcal{K}^{\hat{2}} - \mathbb{S}\mathcal{D}^{\hat{2}}]$ & $0$ & $is P_{2}$ & $0$ & $0$ & $2iJ^{3}$ & $-2i[cD_{\hat{+}} + sD_{\hat{-}}]$ & $0$ & $i P_{\hat{+}}$ & $0$ & $i P_{\hat{-}}$ & $i P_{1}$ \\
\hline
\rule{0pt}{16pt} $\mathfrak{K}_{1}$ & $0$ & $2i\mathcal{D}^{\hat{1}}$ & $-ic\mathfrak{K}_{1}$ & $0$ & $2i\mathcal{K}^{\hat{1}}$ & $-is\mathfrak{K}_{1}$ & $2i[cD_{\hat{+}} + sD_{\hat{-}}]$ & $-2iJ^{3}$ & $0$ & $0$ & $i[\mathbb{C}\mathfrak{K}_{\hat{+}} + \mathbb{S}\mathfrak{K}_{\hat{-}}]$ & $0$ & $i[\mathbb{S}\mathfrak{K}_{\hat{+}} - \mathbb{C}\mathfrak{K}_{\hat{-}}]$ & $0$ & $-i\mathfrak{K}_{2}$ \\
\hline
\rule{0pt}{16pt} $\mathfrak{K}_{2}$ & $0$ & $2i\mathcal{D}^{\hat{2}}$ & $-ic\mathfrak{K}_{2}$ & $0$ & $2i\mathcal{K}^{\hat{2}}$ & $-is\mathfrak{K}_{2}$ & $2iJ^{3}$ & $2i[cD_{\hat{+}} + sD_{\hat{-}}]$ & $0$ & $0$ & $0$ & $i[\mathbb{C}\mathfrak{K}_{\hat{+}} + \mathbb{S}\mathfrak{K}_{\hat{-}}]$ & $0$ & $i[\mathbb{S}\mathfrak{K}_{\hat{+}} - \mathbb{C}\mathfrak{K}_{\hat{-}}]$ & $i\mathfrak{K}_{1}$ \\
\hline
\rule{0pt}{16pt} $\mathcal{D}^{\hat{1}}$ & $-i\mathfrak{K}_{1}$ & $-i\mathbb{C} P_{1}$ & $-is[\mathbb{C}\mathcal{K}^{\hat{1}} - \mathbb{S}\mathcal{D}^{\hat{1}}]$ & $0$ & $-i\mathbb{S} P_{1}$ & $ic[\mathbb{C}\mathcal{K}^{\hat{1}} - \mathbb{S}\mathcal{D}^{\hat{1}}]$ & $-i P_{\hat{+}}$ & $0$ & $-i[\mathbb{C}\mathfrak{K}_{\hat{+}} + \mathbb{S}\mathfrak{K}_{\hat{-}}]$ & $0$ & $0$ & $-i\mathbb{C} J^{3}$ & $-i[sD_{\hat{+}} - cD_{\hat{-}}]$ & $-i\mathbb{S} J^{3}$ & $-i\mathcal{D}^{\hat{2}}$ \\
\hline
\rule{0pt}{16pt} $\mathcal{D}^{\hat{2}}$ & $-i\mathfrak{K}_{2}$ & $-i\mathbb{C} P_{2}$ & $-is[\mathbb{C}\mathcal{K}^{\hat{2}} - \mathbb{S}\mathcal{D}^{\hat{2}}]$ & $0$ & $-i\mathbb{S} P_{2}$ & $ic[\mathbb{C}\mathcal{K}^{\hat{2}} - \mathbb{S}\mathcal{D}^{\hat{2}}]$ & $0$ & $-i P_{\hat{+}}$ & $0$ & $-i[\mathbb{C}\mathfrak{K}_{\hat{+}} + \mathbb{S}\mathfrak{K}_{\hat{-}}]$ & $i\mathbb{C} J^{3}$ & $0$ & $i\mathbb{S} J^{3}$ & $-i[sD_{\hat{+}} - cD_{\hat{-}}]$ & $i\mathcal{D}^{\hat{1}}$ \\
\hline
\rule{0pt}{16pt} $\mathcal{K}^{\hat{1}}$ & $0$ & $-i\mathbb{S} P_{1}$ & $-is[\mathbb{S}\mathcal{K}^{\hat{1}} + \mathbb{C}\mathcal{D}^{\hat{1}}]$ & $-i\mathfrak{K}_{1}$ & $i\mathbb{C} P_{1}$ & $ic[\mathbb{S}\mathcal{K}^{\hat{1}} + \mathbb{C}\mathcal{D}^{\hat{1}}]$ & $-i P_{\hat{-}}$ & $0$ & $-i[\mathbb{S}\mathfrak{K}_{\hat{+}} - \mathbb{C}\mathfrak{K}_{\hat{-}}]$ & $0$ & $i[sD_{\hat{+}} - cD_{\hat{-}}]$ & $-i\mathbb{S} J^{3}$ & $0$ & $i\mathbb{C} J^{3}$ & $-i\mathcal{K}^{\hat{2}}$ \\
\hline
\rule{0pt}{16pt} $\mathcal{K}^{\hat{2}}$ & $0$ & $-i\mathbb{S} P_{2}$ & $-is[\mathbb{S}\mathcal{K}^{\hat{2}} + \mathbb{C}\mathcal{D}^{\hat{2}}]$ & $-i\mathfrak{K}_{2}$ & $i\mathbb{C} P_{2}$ & $ic[\mathbb{S}\mathcal{K}^{\hat{2}} + \mathbb{C}\mathcal{D}^{\hat{2}}]$ & $0$ & $-i P_{\hat{-}}$ & $0$ & $-i[\mathbb{S}\mathfrak{K}_{\hat{+}} - \mathbb{C}\mathfrak{K}_{\hat{-}}]$ & $i\mathbb{S} J^{3}$ & $i[sD_{\hat{+}} - cD_{\hat{-}}]$ & $-i\mathbb{C} J^{3}$ & $0$ & $i\mathcal{K}^{\hat{1}}$ \\
\hline
\rule{0pt}{16pt} $J^{3}$ & $0$ & $0$ & $0$ & $0$ & $0$ & $0$ & $i P_{2}$ & $-i P_{1}$ & $i\mathfrak{K}_{2}$ & $-i\mathfrak{K}_{1}$ & $i\mathcal{D}^{\hat{2}}$ & $-i\mathcal{D}^{\hat{1}}$ & $i\mathcal{K}^{\hat{2}}$ & $-i\mathcal{K}^{\hat{1}}$ & $0$ \\
\hline
\end{tabular}}
\label{tableinterpolation}
\end{minipage}

\newpage

\noindent
\begin{minipage}[t]{\textwidth}
\centering
\renewcommand{\arraystretch}{1.5}
\adjustbox{max width=\textwidth, max totalheight=0.85\textheight}{%
\begin{tabular}{ |c||c|c|c|c|c|c|c|c|c|c|c|c|c|c|c| }
\hline
\rule{0pt}{16pt} & $\mathfrak{K}_{+}$ & $P_{+}$ & $D_{+}$ & $\mathfrak{K}_{-}$ & $P_{-}$ & $D_{-}$ & $P_{1}$ & $P_{2}$ & $\mathfrak{K}_{1}$ & $\mathfrak{K}_{2}$ & $F^{1}$ & $F^{2}$ & $E^{1}$ & $E^{2}$ & $J^{3}$ \\
\hline
\hline
\rule{0pt}{16pt} $\mathfrak{K}_{+}$ & 0 & $-2i\sqrt{2}D_{+}$ & $-i\sqrt{2}\mathfrak{K}_{+}$ & 0 & 0 & 0 & $2iE^{1}$ & $2iE^{2}$ & 0 & 0 & $-i\mathfrak{K}_{1}$ & $-i\mathfrak{K}_{2}$ & 0 & 0 & 0 \\
 \hline
 \rule{0pt}{16pt} $P_{+}$ & $2i\sqrt{2}D_{+}$ & 0 & $i\sqrt{2}P_{+}$ & 0 & 0 & 0 & 0 & 0 & $2iF^{1}$ & $2iF^{2}$ & 0 & 0 & $-iP_{1}$ & $-iP_{2}$ & 0 \\
 \hline
 \rule{0pt}{16pt} $D_{+}$ & $i\sqrt{2}\mathfrak{K}_{+}$ & $-i\sqrt{2}P_{+}$ & 0 & 0 & 0 & 0 & $-i\frac{1}{\sqrt{2}}P_{1}$ & $-i\frac{1}{\sqrt{2}}P_{2}$ & $i\frac{1}{\sqrt{2}}\mathfrak{K}_{1}$ & $i\frac{1}{\sqrt{2}}\mathfrak{K}_{2}$ & $-i\frac{1}{\sqrt{2}}F^{1}$ & $-i\frac{1}{\sqrt{2}}F^{2}$ & $i\frac{1}{\sqrt{2}}E^{1}$ & $i\frac{1}{\sqrt{2}}E^{2}$ & 0 \\
 \hline
 \rule{0pt}{16pt} $\mathfrak{K}_{-}$ & 0 & 0 & 0 & 0 & $-2i\sqrt{2}D_{-}$ & $-i\sqrt{2}\mathfrak{K}_{-}$ & $2iF^{1}$ & $2iF^{2}$ & 0 & 0 & 0 & 0 & $-i\mathfrak{K}_{1}$ & $-i\mathfrak{K}_{2}$ & 0 \\
 \hline
 \rule{0pt}{16pt} $P_{-}$ & 0 & 0 & 0 & $2i\sqrt{2}D_{-}$ & 0 & $i\sqrt{2}P_{-}$ & 0 & 0 & $2iE^{1}$ & $2iE^{2}$ & $-iP_{1}$ & $-iP_{2}$ & 0 & 0 & 0 \\
 \hline
 \rule{0pt}{16pt} $D_{-}$ & 0 & 0 & 0 & $i\sqrt{2}\mathfrak{K}_{-}$ & $-i\sqrt{2}P_{-}$ & 0 & $-i\frac{1}{\sqrt{2}}P_{1}$ & $-i\frac{1}{\sqrt{2}}P_{2}$ & $i\frac{1}{\sqrt{2}}\mathfrak{K}_{1}$ & $i\frac{1}{\sqrt{2}}\mathfrak{K}_{2}$ & $i\frac{1}{\sqrt{2}}F^{1}$ & $i\frac{1}{\sqrt{2}}F^{2}$ & $-i\frac{1}{\sqrt{2}}E^{1}$ & $-i\frac{1}{\sqrt{2}}E^{2}$ & 0 \\
 \hline
 \rule{0pt}{16pt} $P_{1}$ & $-2iE^{1}$ & 0 & $i\frac{1}{\sqrt{2}}P_{1}$ & $-2iF^{1}$ & 0 & $i\frac{1}{\sqrt{2}}P_{1}$ & 0 & 0 & $-i\sqrt{2}(D_{+}+D_{-})$ & $-2iJ^{3}$ & $-iP_{+}$ & 0 & $-iP_{-}$ & 0 & $-iP_{2}$ \\
 \hline
 \rule{0pt}{16pt} $P_{2}$ & $-2iE^{2}$ & 0 & $i\frac{1}{\sqrt{2}}P_{2}$ & $-2iF^{2}$ & 0 & $i\frac{1}{\sqrt{2}}P_{2}$ & 0 & 0 & $2iJ^{3}$ & $-i\sqrt{2}(D_{+}+D_{-})$ & 0 & $-iP_{+}$ & 0 & $-iP_{-}$ & $iP_{1}$ \\
 \hline
 \rule{0pt}{16pt} $\mathfrak{K}_{1}$ & 0 & $-2iF^{1}$ & $-i\frac{1}{\sqrt{2}}\mathfrak{K}_{1}$ & 0 & $-2iE^{1}$ & $-i\frac{1}{\sqrt{2}}\mathfrak{K}_{1}$ & $i\sqrt{2}(D_{+}+D_{-})$ & $-2iJ^{3}$ & 0 & 0 & $-i\mathfrak{K}_{-}$ & 0 & $-i\mathfrak{K}_{+}$ & 0 & $-i\mathfrak{K}_{2}$ \\
 \hline
 \rule{0pt}{16pt} $\mathfrak{K}_{2}$ & 0 & $-2iF^{2}$ & $-i\frac{1}{\sqrt{2}}\mathfrak{K}_{2}$ & 0 & $-2iE^{2}$ & $-i\frac{1}{\sqrt{2}}\mathfrak{K}_{2}$ & $2iJ^{3}$ & $i\sqrt{2}(D_{+}+D_{-})$ & 0 & 0 & 0 & $-i\mathfrak{K}_{-}$ & 0 & $-i\mathfrak{K}_{+}$ & $i\mathfrak{K}_{1}$ \\
 \hline
 \rule{0pt}{16pt} $F^{1}$ & $i\mathfrak{K}_{1}$ & 0 & $i\frac{1}{\sqrt{2}}F^{1}$ & 0 & $iP_{1}$ & $-i\frac{1}{\sqrt{2}}F^{1}$ & $iP_{+}$ & 0 & $i\mathfrak{K}_{-}$ & 0 & 0 & 0 & $-i\frac{1}{\sqrt{2}}(D_{+}-D_{-})$ & $-iJ^{3}$ & $-iF^{2}$\\
 \hline
 \rule{0pt}{16pt} $F^{2}$ & $i\mathfrak{K}_{2}$ & 0 & $i\frac{1}{\sqrt{2}}F^{2}$ & 0 & $iP_{2}$ & $-i\frac{1}{\sqrt{2}}F^{2}$ & 0 & $iP_{+}$ & 0 & $i\mathfrak{K}_{-}$ & 0 & 0 & $iJ^{3}$ & $-i\frac{1}{\sqrt{2}}(D_{+}-D_{-})$ & $iF^{1}$\\
 \hline
 \rule{0pt}{16pt} $E^{1}$ & 0 & $iP_{1}$ & $-i\frac{1}{\sqrt{2}}E^{1}$ & $i\mathfrak{K}_{1}$ & 0 & $i\frac{1}{\sqrt{2}}E^{1}$ & $iP_{-}$ & 0 & $i\mathfrak{K}_{+}$ & 0 & $i\frac{1}{\sqrt{2}}(D_{+}-D_{-})$ & $-iJ^{3}$ & 0 & 0 & $-iE^{2}$ \\
 \hline
 \rule{0pt}{16pt} $E^{2}$ & 0 & $iP_{2}$ & $-i\frac{1}{\sqrt{2}}E^{2}$ & $i\mathfrak{K}_{2}$ & 0 & $i\frac{1}{\sqrt{2}}E^{2}$ & 0 & $iP_{-}$ & 0 & $i\mathfrak{K}_{+}$ & $iJ^{3}$ & $i\frac{1}{\sqrt{2}}(D_{+}-D_{-})$ & 0 & 0 & $iE^{1}$ \\
 \hline
 \rule{0pt}{16pt} $J^{3}$ & 0 & 0 & 0 & 0 & 0 & 0 & $iP_{2}$ & $-iP_{1}$ & $i\mathfrak{K}_{2}$ & $-i\mathfrak{K}_{1}$ & $iF^{2}$ & $-iF^{1}$ & $iE^{2}$ & $-iE^{1}$ & 0 \\
 \hline
\end{tabular}}
\captionof{table}{Conformal algebra $(3+1)$ in the LFD limit ($\delta \to \pi/4$)}
\label{tableLFD}
\end{minipage}
\endgroup
\finishlandscape

\section{\texorpdfstring{$6\times6$}{TEXT} Projective Spacetime Representations}
\label{sec:projectivespace}
To characterize the individual properties of the 15 conformal generators, namely, which generators are kinematic or dynamic, we may correspond the projective spacetime representations 
given by Eq.(\ref{Jab}) with
the interpolating $(3+1)$ dimensional conformal algebra
discussed in the previous sections. 
Kinematic generators leave the time defined in the given form of the dynamics invariant.
Consequently, the individual time-ordered amplitudes in the respective time-ordered processes are invariant under the transformations provided by kinematic generators \cite{Ji2001, Ravikumar:2021xck}.
Determining which generators are kinematic in each form of dynamics is useful, as maximizing the number of kinematic generators correspondingly saves dynamical effort when solving relativistic quantum field theories, such as QED and QCD~ \cite{ji2023relativistic}. As discussed in earlier works~\cite{Ji-Hari-2026-PRD, Ji2001, Ji2025, Ji2015EM, Ji2021QCD,THOOFT1974461}, 
LFD saves dynamical effort in solving relativistic quantum field theoretic problems due to the character change of the longitudinal boost operator $K_3$ from dynamic in IFD to kinematic in LFD. 

 The conformal algebra given by Eq.\eqref{simplesrint} in the interpolating form of projective spacetime implies that $J_{\hat{a}\hat{b}}$ can be written as
\begin{align}
    J_{\hat{a}\hat{b}}=i(X_{\hat{a}}\partial_{\hat{b}}-X_{\hat{b}}\partial_{\hat{a}}),
\end{align}
where $\hat{a},\hat{b}\in\{-2,-1,\hat{+},1,2,\hat{-}\}$, $\partial_{\hat{a}}\equiv\frac{\partial}{\partial X^{\hat{a}}}$, and $X_{\hat{a}}$ is a six-dimensional projective embedding vector for the physical $(3+1)$-dimensional spacetime. With the condition that its lightcone 
$X^{\hat{a}}X_{\hat{a}}=0$
is preserved under the transformations generated by $J_{\hat{a}\hat{b}}$, we write the infinitesimal conformal transformations as
\begin{align}
R^{\hat{a}}_{~\hat{b}}=g^{\hat{a}}_{~\hat{b}}-\frac{i}{2}\omega^{\hat{c}\hat{d}}(J_{\hat{c}\hat{d}})^{\hat{a}}_{~\hat{b}},\label{R_conformalTransformations}
\end{align}
where $\omega^{\hat{a}\hat{b}}=-\omega^{\hat{b}\hat{a}}$. Here, we may regard $R^{\hat{a}}_{~\hat{b}}$ as the expansion of $e^{-\frac{i}{2}\omega^{\hat{c}\hat{d}}(J_{\hat{c}\hat{d}})^{\hat{a}}_{~\hat{b}}}$ up to the first order in $\omega$, as it can be seen in the expansion given by 
$e^{-\frac{i}{2}\omega^{\hat{c}\hat{d}}(J_{\hat{c}\hat{d}})^{\hat{a}}_{~\hat{b}}} \approx  g^{\hat{a}}_{~\hat{b}}-\frac{i}{2}\omega^{\hat{c}\hat{d}}(J_{\hat{c}\hat{d}})^{\hat{a}}_{~\hat{b}}+\mathcal{O}(\omega^2)$.
The generator representation $(J_{\hat{c}\hat{d}})^{\hat{a}}_{~\hat{b}}$ can be obtained by
\begin{align}
    (J_{\hat{c}\hat{d}})^{\hat{a}}_{~\hat{b}}=(J_{\hat{c}\hat{d}})^{\hat{a}\hat{f}}g_{\hat{f}\hat{b}}~~,
\end{align}
where $(J_{{\hat{c}}{\hat{d}}})^{{\hat{a}}{\hat{b}}}=i(g_{\hat{c}}^{~~\hat{a}} g_{\hat{d}}^{~~\hat{b}}-g_{\hat{c}}^{~~\hat{b}} g_{\hat{d}}^{~~\hat{a}})$. The representation matrices of the conformal generators are defined by taking the first index as a superscript and the second as a subscript:

\begin{eqnarray}
    (D_{\hat{+}})^{\hat{a}}_{~\hat{b}}&&\equiv -\cos{\delta}(J_{-2-1})^{\hat{a}}_{~\hat{b}}-\sin{\delta}(J_{\hat{+}\hat{-}})^{\hat{a}}_{~\hat{b}}~;\nonumber\\
    (D_{\hat{-}})^{\hat{a}}_{~\hat{b}}&&\equiv -\sin{\delta}(J_{-2-1})^{\hat{a}}_{~\hat{b}}+\cos{\delta}(J_{\hat{+}\hat{-}})^{\hat{a}}_{~\hat{b}}~;\nonumber\\
    \frac{-(\mathfrak{K}^{\hat{\mu}})^{\hat{a}}_{~\hat{b}}}{\sqrt{2}}&&\equiv (J_{-2\hat{\mu}})^{\hat{a}}_{~\hat{b}}~;\nonumber\\
    \frac{(P_{\hat{\mu}})^{\hat{a}}_{~\hat{b}}}{\sqrt{2}}&&\equiv (J_{-1\hat{\mu}})^{\hat{a}}_{~\hat{b}}~;\nonumber\\
    (\mathcal{D}^{\hat{i}})^{\hat{a}}_{~\hat{b}}&&\equiv (J_{\hat{+}i})^{\hat{a}}_{~\hat{b}}~;\nonumber\\(\mathcal{K}^{\hat{i}})^{\hat{a}}_{~\hat{b}}&&\equiv (J_{\hat{-}i})^{\hat{a}}_{~\hat{b}}~;\nonumber\\
    (J^3)^{\hat{a}}_{~\hat{b}}&&\equiv (J_{12})^{\hat{a}}_{~\hat{b}}~.
    \label{implicit6x6}
\end{eqnarray}

The explicit $6\times6$ matrix representations are then given by
\begin{align}
    D_{\hat{+}}&=\begin{pmatrix}
        -i\cos{(\delta)}&0&0&0&0&0\\
        0&i\cos{(\delta)}&0&0&0&0\\
        0&0&i\mathbb{S}\sin{(\delta)}&0&0&-i\mathbb{C}\sin{(\delta)}\\
        0&0&0&0&0&0\\
        0&0&0&0&0&0\\
        0&0&-i\mathbb{C}\sin{(\delta)}&0&0&-i\mathbb{S}\sin{(\delta)}
    \end{pmatrix};\nonumber\\
    D_{\hat{-}}&=\begin{pmatrix}
        -i\sin{(\delta)}&0&0&0&0&0\\
        0&i\sin{(\delta)}&0&0&0&0\\
        0&0&-i\mathbb{S}\cos{(\delta)}&0&0&i\mathbb{C}\cos{(\delta)}\\
        0&0&0&0&0&0\\
        0&0&0&0&0&0\\
        0&0&i\mathbb{C}\cos{(\delta)}&0&0&i\mathbb{S}\cos{(\delta)}
    \end{pmatrix};\nonumber
\end{align}
\begin{align}
    \mathfrak{K}^{\hat{+}}&=\sqrt{2}\begin{pmatrix}
        0&0&0&0&0&0\\
        0&0&i&0&0&0\\
        i\mathbb{C}&0&0&0&0&0\\
        0&0&0&0&0&0\\
        0&0&0&0&0&0\\
        i\mathbb{S}&0&0&0&0&0
    \end{pmatrix};
    ~\mathfrak{K}^{\hat{-}}=\sqrt{2}\begin{pmatrix}
        0&0&0&0&0&0\\
        0&0&0&0&0&i\\
        i\mathbb{S}&0&0&0&0&0\\
        0&0&0&0&0&0\\
        0&0&0&0&0&0\\
        -i\mathbb{C}&0&0&0&0&0
    \end{pmatrix};\nonumber
    \end{align}
    \begin{align}
     P_{\hat{+}}&=\sqrt{2}\begin{pmatrix}
        0&0&-i&0&0&0\\
        0&0&0&0&0&0\\
        0&-i\mathbb{C}&0&0&0&0\\
        0&0&0&0&0&0\\
        0&0&0&0&0&0\\
        0&-i\mathbb{S}&0&0&0&0
    \end{pmatrix}~;P_{\hat{-}}=\sqrt{2}\begin{pmatrix}
        0&0&0&0&0&-i\\
        0&0&0&0&0&0\\
        0&-i\mathbb{S}&0&0&0&0\\
        0&0&0&0&0&0\\
        0&0&0&0&0&0\\
        0&i\mathbb{C}&0&0&0&0
    \end{pmatrix}~;\nonumber
    \end{align}
    \begin{align}
        \mathcal{K}^{\hat{1}}=\begin{pmatrix}
        0&0&0&0&0&0\\
        0&0&0&0&0&0\\
        0&0&0&i\mathbb{S}&0&0\\
        0&0&0&0&0&i\\
        0&0&0&0&0&0\\
        0&0&0&-i\mathbb{C}&0&0
    \end{pmatrix}~;~~~~~~\mathcal{K}^{\hat{2}}=\begin{pmatrix}
        0&0&0&0&0&0\\
        0&0&0&0&0&0\\
        0&0&0&0&i\mathbb{S}&0\\
        0&0&0&0&0&0\\
        0&0&0&0&0&i\\
        0&0&0&0&-i\mathbb{C}&0
    \end{pmatrix}~;\nonumber
    \end{align}
    \begin{align}
    \mathcal{D}^{\hat{1}}&=\begin{pmatrix}
        0&0&0&0&0&0\\
        0&0&0&0&0&0\\
        0&0&0&i\mathbb{C}&0&0\\
        0&0&i&0&0&0\\
        0&0&0&0&0&0\\
        0&0&0&i\mathbb{S}&0&0
    \end{pmatrix}~;~~~~~~\mathcal{D}^{\hat{2}}=\begin{pmatrix}
        0&0&0&0&0&0\\
        0&0&0&0&0&0\\
        0&0&0&0&i\mathbb{C}&0\\
        0&0&0&0&0&0\\
        0&0&i&0&0&0\\
        0&0&0&0&i\mathbb{S}&0
    \end{pmatrix}~;\nonumber
    \end{align}
    \begin{align}
        \mathfrak{K}_1=\sqrt{2}\begin{pmatrix}
    0&0&0&0&0&0\\
    0&0&0&i&0&0\\
    0&0&0&0&0&0\\
    -i&0&0&0&0&0\\
    0&0&0&0&0&0\\
    0&0&0&0&0&0
    \end{pmatrix}~;
    \mathfrak{K}_2&=\sqrt{2}\begin{pmatrix}
    0&0&0&0&0&0\\
    0&0&0&0&i&0\\
    0&0&0&0&0&0\\
    0&0&0&0&0&0\\
    -i&0&0&0&0&0\\
    0&0&0&0&0&0
    \end{pmatrix}~;\nonumber
    \end{align}
    \begin{align}
        P_1=\sqrt{2}\begin{pmatrix}
    0&0&0&-i&0&0\\
    0&0&0&0&0&0\\
    0&0&0&0&0&0\\
    0&i&0&0&0&0\\
    0&0&0&0&0&0\\
    0&0&0&0&0&0
    \end{pmatrix}~;~~~~~~
    P_2=\sqrt{2}\begin{pmatrix}
    0&0&0&0&-i&0\\
    0&0&0&0&0&0\\
    0&0&0&0&0&0\\
    0&0&0&0&0&0\\
    0&i&0&0&0&0\\
    0&0&0&0&0&0
    \end{pmatrix}~;\nonumber
    \end{align}
    \begin{align}
        J^3&=\begin{pmatrix}
    0&0&0&0&0&0\\
    0&0&0&0&0&0\\
    0&0&0&0&0&0\\
  0&0&0&0 & -i & 0\\
  0&0&0&i & 0 & 0\\
  0&0&0&0 & 0 & 0
\end{pmatrix}.\label{matrix-rep-conformal-generators}
    \end{align}
These explicit projective spacetime matrices satisfy the full conformal algebra given by Eq.\eqref{simplesrint} in the  interpolating dynamics between IFD and LFD. 

Operating the exponentiated conformal transformations $e^{-\frac{i}{2}\omega^{\hat{c}\hat{d}}(J_{\hat{c}\hat{d}})^{\hat{a}}_{~\hat{b}}}$ on $X_{\hat{b}}$, we obtain the transformed six-dimensional projective embedding vector $X^{\prime}_{\hat{a}}$ as
\begin{align}
\label{exponential-conformal-tranf}
  X^{\prime}_{\hat{a}}&=e^{-\frac{i}{2}\omega^{\hat{c}\hat{d}}(J_{\hat{c}\hat{d}})^{\hat{b}}_{~\hat{a}}}X_{\hat{b}}~,
\end{align}
and the inner product
\begin{align}
{X^{\prime}}_{\hat{a}}{X^{\prime}}^{\hat{a}} &= 
e^{-\frac{i}{2}\omega^{\hat{c}\hat{d}}(J_{\hat{c}\hat{d}})^{\hat{b}}_{~\hat{a}}}X_{\hat{b}}e^{-\frac{i}{2}\omega^{\hat{c}\hat{d}}(J_{\hat{c}\hat{d}})^{\hat{a}}_{~\hat{b}}}X^{\hat{b}}\nonumber\\
&= e^{-\frac{i}{2}\omega^{\hat{c}\hat{d}}(J_{\hat{c}\hat{d}})^{\hat{b}}_{~\hat{a}}}e^{-\frac{i}{2}\omega^{\hat{c}\hat{d}}(J_{\hat{c}\hat{d}})^{\hat{a}}_{~\hat{b}}}X_{\hat{b}}X^{\hat{b}}\nonumber\\
&= e^{-\frac{i}{2}\omega^{\hat{c}\hat{d}}((J_{\hat{c}\hat{d}})^{\hat{b}}_{~\hat{a}}+(J_{\hat{c}\hat{d}})^{\hat{a}}_{~\hat{b}})}X_{\hat{b}}X^{\hat{b}}\nonumber\\
&= X_{\hat{b}}X^{\hat{b}},
\end{align}
where we used 
$(J_{{\hat{c}}{\hat{d}}})^{\hat{a}}_{~\hat{b}}=i(g_{\hat{c}}^{~~\hat{a}} g_{{\hat{d}}{\hat{b}}}-g_{{\hat{c}}{\hat{b}}} g_{\hat{d}}^{~~\hat{a}})$ 
to obtain $(J_{\hat{c}\hat{d}})^{\hat{a}}_{~\hat{b}}=-(J_{\hat{c}\hat{d}})^{\hat{b}}_{~\hat{a}}$.
This shows that the interpolating conformal transformations preserve the angles and the ratios of lengths of two vectors before and after the transformations. Now, we assign the following six-dimensional projective embedding vectors:
\begin{align}
    {X}_{-1}&=\frac{-\lambda}{\sqrt{2}}, \nonumber\\
    {X}_{-2}&=\frac{-\lambda}{\sqrt{2}}(x^{\hat{\mu}}x_{\hat{\mu}}), \nonumber \\
     {X}_{\hat{\mu}}&=\lambda x_{\hat{\mu}},\label{X-mu}
\end{align}
providing the lightcone of the projective spacetime, namely ${X}_{\hat{a}}{X}^{\hat{a}}=0$, as shown explicitly below: 
\begin{align}
    {X}_{\hat{a}}{X}^{\hat{a}}&={X}_{-2}{X}^{-2}+{X}_{-1}{X}^{-1}+{X}_{\hat{\mu}}{X}^{\hat{\mu}} \nonumber \\
    &=-2{X}_{-2}{X}_{-1}+{X}^{\hat{\mu}}{X}_{\hat{\mu}}\nonumber \\
     &=-2\frac{-\lambda}{\sqrt{2}}(x^{\hat{\mu}}x_{\hat{\mu}})\frac{-\lambda}{\sqrt{2}}+\lambda^2{x}^{\hat{\mu}}{x}_{\hat{\mu}}\nonumber \\
     &=-\lambda^2{x}^{\hat{\mu}}{x}_{\hat{\mu}}+\lambda^2{x}^{\hat{\mu}}{x}_{\hat{\mu}}\nonumber \\
     &=0~,
\end{align}
where ${X}^{-1}=-{X}_{-2}$, ${X}^{-2}=-{X}_{-1}$ and $X^{\hat \mu}=g^{\hat{\mu}\hat{\nu}}X_{\hat{\nu}}$ due to the metric
given by Eq.(\ref{metricghat}).
From Eq.\eqref{X-mu}, we can then find 
the dictionary connecting the six-dimensional projective embedding vector $X_{\hat{a}}$ and the physical $(3+1)$-dimensional spacetime $x_{\hat{\mu}}$, which reads
\begin{align}
    x_{\hat{\mu}}&=-\frac{1}{\sqrt{2}}\frac{{X}_{\hat{\mu}}}{ {X}_{-1}}.
\end{align}
Corresponding to $x_{\hat{\mu}}$, we have  
\begin{align}
    x^{\hat{\mu}}&=-\frac{1}{\sqrt{2}}\frac{{X}^{\hat{\mu}}}{ {X}_{-1}}.
\end{align}
Then, the inverse spacetime of $x^{\hat{\mu}}$ is given by $\frac{x_{\hat{\mu}}}{x^2}=-\frac{1}{\sqrt{2}}\frac{{X}_{\hat{\mu}}}{{X}_{-2}}$, with $x^2=\frac{{X}_{-2}}{{X}_{-1}}$. Under the transformation given by Eq.(\ref{exponential-conformal-tranf}), we compute the transformation of interpolating time given by
\begin{align}
    x^{\hat{+}\prime}&=-\frac{1}{\sqrt{2}}\frac{{X}^{\hat{+}\prime}}{ {X}^{\prime}_{-1}},
\end{align}
for each and every interpolating conformal operation according to Eq.(\ref{exponential-conformal-tranf}).
The details of the explicit derivations are presented in Appendix \ref{app:interpolating-time-derivations}.

Summarizing the results of our computation in Table~\ref{tablexprime}, we can determine which generators leave the interpolating time intact. 
As mentioned earlier in the Poincar\'e algebra, among the ten Poincar\'e generators, the six generators ($\mathcal{K}^{\itP{1}}, \mathcal{K}^{\itP{2}}, J^{3}, P_{1}, P_{2}, P_{\mT}$) are kinematic in the sense that the $x^{\pT}=0$ plane remains intact under the transformations generated by them.
Now, in the conformal algebra, we note that the seven generators, including the dilatation generator $D=\left(D_{\hat{+}}c+D_{\hat{-}}s\right)=J_{-1 -2}$, are kinematic, while we added five more generators ($D$, $\mathfrak{K}_{\hat{+}}$, $\mathfrak{K}_{{1}}$, $\mathfrak{K}_{{2}}$, $\mathfrak{K}_{\hat{-}}$) beyond the 
ten Poincar\'e generators
($\mathcal{K}^{\hat{1}}, \mathcal{K}^{\hat{2}}, J^{3}, P^{1}, P^{2}, P_{\mT},  \mathcal{D}^{\hat{1}}, \mathcal{D}^{\hat{2}}, K_3, P_{\pT}$). 
Note here that the dilatation $D$ and longitudinal boost $K_3$ generators are specific combinations of $D_{\hat{+}}$ and $D_{\hat{-}}$ according to Eq.(\ref{mixingDandK3}).
Among the fifteen generators of the conformal algebra, the seven generators ($D = (D_{\hat{+}}c + D_{\hat{-}}s)=J_{-1-2}, \mathcal{K}^{\itP{1}}, \mathcal{K}^{\itP{2}}, J^{3}, P_{1}, P_{2}, P_{\mT}$) are kinematic and 
the other eight generators ($\mathfrak{K}_{\hat{+}}, \mathfrak{K}_{\hat{-}}, P_{\hat{+}}, K_3=\left(D_{\hat{+}}s-D_{\hat{-}}c\right)=-J_{{\hat +}{\hat -}}, \mathfrak{K}_{{1}}, \mathfrak{K}_{{2}},\mathcal{D}^1,
\mathcal{D}^2$) are dynamic. 
While the kinematic generators leave the $x^{\pT}=0$ plane intact under the corresponding transformations, the dynamic generators transform to $x'^{\pT}$ different from $x^{\pT}$ as summarized in Table~\ref{tablexprime}. 
Taking the respective limit of the interpolating parameter $\delta$, namely,  
$\delta \to 0$ and $\delta \to \pi/4$, we find the corresponding number of kinematic generators in the IFD and the LFD, respectively, as summarized below.
\begin{table}[h!]
        \resizebox{\textwidth}{!}{
        \begin{tabular}{|c||c|c|}
        \hline
             \rule{0pt}{16pt} Generators & $x^{\hat{+}\prime}$\\
             \hline
            \rule{0pt}{16pt} $\mathfrak{K}^{\hat{+}}$ & $x^{\hat{+}\prime}=\frac{x^{\hat{+}}-b^{\hat{+}}(x)^2}{1-2\mathbb{C}b^{\hat{+}}x^{\hat{+}}-2\mathbb{S}b^{\hat{+}}x^{\hat{-}}+\mathbb{C}(b^{\hat{+}})^2(x)^2}$\\
             \hline
            \rule{0pt}{16pt} $\mathfrak{K}^{\hat{-}}$ & $x^{\hat{+}\prime}=\frac{x^{\hat{+}}}{1+2\mathbb{C}b^{\hat{-}}x^{\hat{-}}-2\mathbb{S}b^{\hat{-}}x^{\hat{+}}-\mathbb{C}(b^{\hat{-}})^2(x)^2}$ \\
             \hline
             \rule{0pt}{16pt} $P_{\hat{+}}$& $x^{\hat{+}\prime}=x^{\hat{+}}+a^{\hat{+}} $\\
             \hline
             \rule{0pt}{16pt} $P_{\hat{-}}$& $x^{\hat{+}\prime}=x^{\hat{+}}$  \\
             \hline
             \rule{0pt}{16pt} $D_{\hat{+}}$ & $x^{\hat{+}\prime} = e^{-\alpha_{\hat{+}}\sin{(\delta)}}\left[\left(\cosh[\alpha_{\hat{+}}\cos{(\delta)}]-\mathbb{S}\sinh[\alpha_{\hat{+}}\cos{(\delta)}]\right)x^{\hat{+}}+\mathbb{C}\sinh[\alpha_{\hat{+}}\cos{(\delta)}] x^{\hat{-}}\right]$ \\
             \hline
             \rule{0pt}{16pt} $D_{\hat{-}}$ & $x^{\hat{+}\prime} = e^{-\alpha_{\hat{-}}\cos{(\delta)}}\left[\left(\cosh[\alpha_{\hat{-}}\sin{(\delta)}]+\mathbb{S} \sinh[\alpha_{\hat{-}}\sin{(\delta)}]\right)x^{\hat{+}} -\mathbb{C} \sinh[\alpha_{\hat{-}}\sin{(\delta)}]x^{\hat{-}}\right]$ \\
             \hline
             \rule{0pt}{16pt} $J^{3}$ & $ x^{\hat{+}\prime}=x^{\hat{+}}$ \\
             \hline
             \rule{0pt}{16pt} $\mathfrak{K}_{1}$ & $x^{\hat{+}\prime}=\frac{x^{{\hat{+}}}}{1-2b^{1}x_{1}-(b^{1})^{2}(x)^2}$ \\
             \hline
             \rule{0pt}{16pt} $\mathfrak{K}_{2}$ &  $x^{\hat{+}\prime}=\frac{x^{{\hat{+}}}}{1-2b^{2}x_{2}-(b^{2})^{2}(x)^2}$ \\
             \hline
              \rule{0pt}{16pt} $P_{1}$ & $ x^{\hat{+}\prime}=x^{\hat{+}}$  \\
              \hline
              \rule{0pt}{16pt} $P_{2}$ & $ x^{\hat{+}\prime}=x^{\hat{+}}$  \\
              \hline
               \rule{0pt}{16pt} $\mathcal{D}^{\hat{1}}$ &  $ x^{\hat{+}\prime}=\cosh{(\sqrt{\mathbb{C}}\eta^{1})}x^{\hat{+}}+\frac{\mathbb{S}}{\mathbb{C}}\left(\cosh{(\sqrt{\mathbb{C}}\eta^{1})}-1\right)x^{\hat{-}}+\frac{1}{\sqrt{\mathbb{C}}}\sinh{(\sqrt{\mathbb{C}}\eta^{1})}x_{{1}}$ \\
               \hline
               \rule{0pt}{16pt} $\mathcal{D}^{\hat{2}}$ & $ x^{\hat{+}\prime}=\cosh{(\sqrt{\mathbb{C}}\eta^{2})}x^{\hat{+}}+\frac{\mathbb{S}}{\mathbb{C}}\left(\cosh{(\sqrt{\mathbb{C}}\eta^{2})}-1\right)x^{\hat{-}}+\frac{1}{\sqrt{\mathbb{C}}}\sinh{(\sqrt{\mathbb{C}}\eta^{2})}x_{{2}}$ \\
               \hline
               \rule{0pt}{16pt} $\mathcal{K}^{\hat{1}}$ & $ x^{\hat{+}\prime}=x^{\hat{+}}$ \\
               \hline
               \rule{0pt}{16pt} $\mathcal{K}^{\hat{2}}$ & $ x^{\hat{+}\prime}=x^{\hat{+}}$ \\
               \hline
        \end{tabular}}
        \caption{Transformation of interpolating time under each conformal generator in $(3+1)$ dimensions. Each row denotes a one-parameter transformation $X'=e^{-i\lambda_G G}X$, with the parameter associated with the displayed generator $G$: $b^{\hat{\pm}}$, $b^1$, and $b^2$ for special conformal transformations; $a^{\hat{\pm}}$, $a^1$, and $a^2$ for translations; $\alpha_{\hat{\pm}}$ for $D_{\hat{\pm}}$; $\eta^1$ and $\eta^2$ for $\mathcal{D}^{\hat{1}}$ and $\mathcal{D}^{\hat{2}}$; $\phi^1$ and $\phi^2$ for $\mathcal{K}^{\hat{1}}$ and $\mathcal{K}^{\hat{2}}$; and $\theta$ for $J^3$. The interpolating parameter between IFD and LFD is $\delta$.}
        \label{tablexprime}
    \end{table}
\FloatBarrier

\newcommand{\IFDTimeTable}{%
    \begin{table}[!htbp]
      \centering
      \small
        \begin{tabular}{|c||c|}
        \hline
             \rule{0pt}{16pt} Generators & $x^{{0}\prime}$\\
             \hline
             \hline
            \rule{0pt}{16pt} $\mathfrak{K}_{{0}}$ & $x^{{0}\prime}=\frac{x^{{0}}-b^{{0}}(x)^2}{1-2b^{{0}}x^{{0}}+(b^{{0}})^2(x)^2}$\\
             \hline
            \rule{0pt}{16pt} $-\mathfrak{K}_{{3}}$ & $x^{{0}\prime}=\frac{x^{{0}}}{1+2b^{{3}}x^{{3}}+(b^{{3}})^2(x)^2}$ \\
             \hline
             \rule{0pt}{16pt} $P_{{0}}$& $x^{{0}\prime}=x^{{0}}+a^{{0}} $\\
             \hline
             \rule{0pt}{16pt} $-P_{{3}}$& $x^{{0}\prime}=x^{{0}}$  \\
             \hline
             \rule{0pt}{16pt} $D$ & $x^{{0}\prime} = e^{-\alpha}x^{0}$ \\
             \hline
             \rule{0pt}{16pt} $-K_{3}$ & $x^{{0}\prime} = \cosh[\alpha]x^{0}-\sinh[\alpha] x^{3}$ \\
             \hline
             \rule{0pt}{16pt} $J^{3}$ & $ x^{{0}\prime}=x^{{0}}$ \\
             \hline
             \rule{0pt}{16pt} $\mathfrak{K}_{1}$ & $x^{{0}\prime}=\frac{x^{{{0}}}}{1-2b^{1}x_{1}-(b^{1})^{2}(x)^2}$ \\
             \hline
             \rule{0pt}{16pt} $\mathfrak{K}_{2}$ &  $x^{{0}\prime}=\frac{x^{{{0}}}}{1-2b^{2}x_{2}-(b^{2})^{2}(x)^2}$ \\
             \hline
              \rule{0pt}{16pt} $P_{1}$ & $ x^{{0}\prime}=x^{{0}}$  \\
              \hline
              \rule{0pt}{16pt} $P_{2}$ & $ x^{{0}\prime}=x^{{0}}$  \\
              \hline
               \rule{0pt}{16pt} $-K^{1}$ &  $ x^{{0}\prime}=\cosh{(\eta^{1})}x^{{0}}+\sinh{(\eta^{1})}x_{{1}}$ \\
               \hline
               \rule{0pt}{16pt} $-K^{2}$ & $ x^{{0}\prime}=\cosh{(\eta^{2})}x^{{0}}+\sinh{(\eta^{2})}x_{{2}}$ \\
               \hline
               \rule{0pt}{16pt} $-J^{2}$ & $ x^{{0}\prime}=x^{{0}}$ \\
               \hline
               \rule{0pt}{16pt} $J^{1}$ & $ x^{{0}\prime}=x^{{0}}$ \\
               \hline
        \end{tabular}
        \caption{Transformation of the IFD time in $(3+1)$d}
        \label{tablex0prime}
    \end{table}%
}

\newcommand{\LFDTimeTable}{%
    \begin{table}[!htbp]
        \centering
        \small
        \begin{tabular}{|c||c|}
        \hline
             \rule{0pt}{16pt} Generators & $x^{{+}\prime}$\\
             \hline
             \hline
            \rule{0pt}{16pt} $\mathfrak{K}_{{-}}$ & $x^{{+}\prime}=\frac{x^{{+}}-b^{{+}}(2x^{+}x^{-}-\vec{x}^{\perp}\vec{x}^{\perp})}{1-2b^{{+}}x^{{-}}}$\\
             \hline
            \rule{0pt}{16pt} $\mathfrak{K}_{{+}}$ & $x^{{+}\prime}=\frac{x^{{+}}}{1-2b^{{-}}x^{{+}}}$ \\
             \hline
             \rule{0pt}{16pt} $P_{{+}}$& $x^{{+}\prime}=x^{{+}}+a^{{+}} $\\
             \hline
             \rule{0pt}{16pt} $P_{{-}}$& $x^{{+}\prime}=x^{{+}}$  \\
             \hline
             \rule{0pt}{16pt} $D_{{+}}$ & $x^{{+}\prime} = e^{-\sqrt{2}\alpha}x^{{+}}$ \\
             \hline
             \rule{0pt}{16pt} $D_{{-}}$ & $x^{{+}\prime} = x^{{+}}$ \\
             \hline
             \rule{0pt}{16pt} $J^{3}$ & $ x^{{+}\prime}=x^{{+}}$ \\
             \hline
             \rule{0pt}{16pt} $\mathfrak{K}_{1}$ & $x^{{+}\prime}=\frac{x^{{{+}}}}{1-2b^{1}x_{1}-(b^{1})^{2}(x)^2}$ \\
             \hline
             \rule{0pt}{16pt} $\mathfrak{K}_{2}$ &  $x^{{+}\prime}=\frac{x^{{{+}}}}{1-2b^{2}x_{2}-(b^{2})^{2}(x)^2}$ \\
             \hline
              \rule{0pt}{16pt} $P_{1}$ & $ x^{{+}\prime}=x^{{+}}$  \\
              \hline
              \rule{0pt}{16pt} $P_{2}$ & $ x^{{+}\prime}=x^{{+}}$  \\
              \hline
               \rule{0pt}{16pt} $-F^{1}$ &  $x^{{+}\prime}=x^{{+}}+\frac{(\eta^{1})^2}{2}x^{{-}}+\eta^{1}x_{{1}}$ \\
               \hline
               \rule{0pt}{16pt} $-F^{2}$ & $ x^{{+}\prime}=x^{{+}}+\frac{(\eta^{2})^2}{2}x^{{-}}+\eta^{2}x_{{2}}$ \\
               \hline
               \rule{0pt}{16pt} $-E^{1}$ & $ x^{{+}\prime}=x^{{+}}$ \\
               \hline
               \rule{0pt}{16pt}$-E^{2}$ & $ x^{{+}\prime}=x^{{+}}$ \\
               \hline
        \end{tabular}
        \caption{Transformation of LFD time in $(3+1)$d}
        \label{tablexplusprime}
    \end{table}%
}

\newcommand{\KinematicGeneratorTable}{%
    \begin{table}[!htbp]
    \resizebox{\textwidth}{!}{%
    \begin{tabular}{lcc}
	Interpolation angle & Kinematic & Dynamic \\
	\hline
	\rule{0pt}{3ex} $\delta=0$ & $\mathcal{K}^{\hat{1}}=-J^{2},~ \mathcal{K}^{\hat{2}}=J^{1},~ J^{3}, P^{1}, P^{2}, P^{3}$, $D$ & $\mathcal{D}^{\hat{1}}=-K^{1},~ \mathcal{D}^{\hat{2}}=-K^{2},~ K^{3}, P^{0}$, $\mathfrak{K}_{{0}}$, $\mathfrak{K}_{{1}}$, $\mathfrak{K}_{{2}}$, $\mathfrak{K}_{{3}}$\\
	$0<\delta<\pi/4$ &  $\mathcal{K}^{\hat{1}}, \mathcal{K}^{\hat{2}}, J^{3}, P^{1}, P^{2}$, $P_{\mT}$, $D=\left(D_{\hat{+}}c+D_{\hat{-}}s\right)$ &  $\mathcal{D}^{\hat{1}}, \mathcal{D}^{\hat{2}}$, $K_3=\left(D_{\hat{+}}s-D_{\hat{-}}c\right)$,  $P_{\pT}$, $\mathfrak{K}_{\hat{+}}$, $\mathfrak{K}_{1}$, $\mathfrak{K}_{2}$, $\mathfrak{K}_{\hat{-}}$\\
	$\delta=\pi/4$ & $\mathcal{K}^{\hat{1}}=-E^{1},~ \mathcal{K}^{\hat{2}}=-E^{2},~ J^{3}, P^{1}, P^{2}$, $P_{-}$, $D_{+}$, $D_{-}$& $\mathcal{D}^{\hat{1}}=-F^{1},~ \mathcal{D}^{\hat{2}}=-F^{2},~ P_{+}$, $\mathfrak{K}_{{+}}$, $\mathfrak{K}_{{-}}$, $\mathfrak{K}_{{1}}$, $\mathfrak{K}_{{2}}$\\
      \end{tabular}%
    }
    \caption{\label{tab:Kinematic_and_dynamic_generators_for_different_interoplation_angles_conformal1+1}Kinematic and dynamic conformal generators for different interpolation angles in $(3+1)$ dimensions. Here, kinematic means that the interpolating time is unchanged up to a scale factor independent of the spacetime coordinates.}
\end{table}%
}

In the limit to $\delta=0$, i.e., the IFD ($\mathbb{C}=1;~\mathbb{S}=0$), we find the results shown in Table~\ref{tablex0prime} for the transformed $x^{0\prime}$ 
for every $(3+1)$d conformal generator. Consistent with Table~\ref{tablexprime}, the seven kinematic generators are $\mathcal{K}^{\hat{1}}=-J^{2}$, $\mathcal{K}^{\hat{2}}=J^{1}$, $J^{3}$, $P^{1}$, $P^{2}$, $P^{3}=-P_{3}$, and $D$.

\IFDTimeTable
\FloatBarrier

In the limit to $\delta=\pi/4$, i.e., the LFD ($\mathbb{C}=0;~\mathbb{S}=1$), we find the results shown in Table~\ref{tablexplusprime} for the transformed $x^{+\prime}$ for each $(3+1)$d conformal generator. The eight kinematic generators are $\mathcal{K}^{\hat{1}}=-E^{1}$, $\mathcal{K}^{\hat{2}}=-E^{2}$, $J^{3}$, $P^{1}$, $P^{2}$, $P_{-}=\frac{P_{0}-P_{3}}{\sqrt{2}}$, $D_{+}=\frac{(D+K_3)}{\sqrt{2}}$, and $D_{-}=\frac{(D-K_3)}{\sqrt{2}}$. Equivalently, the latter two generators span the kinematic directions $D$ and $K_3$, where $K_3=\left(D_{\hat{+}}s-D_{\hat{-}}c\right)=J_{\hat{-}\hat{+}}$ at $\delta=\pi/4$.

\LFDTimeTable
\FloatBarrier

Finally, we may contrast the behavior of the special conformal transformations (SCT) between IFD and LFD in $(3+1)$ dimensions with the $(1+1)$ dimensional case. In $(1+1)$ dimensions, the transverse coordinates vanish ($\vec{x}^{\perp} = 0$), and the transformation under $\mathfrak{K}_{-}$ simplifies to $x^{+\prime} = x^{+}$, making it an additional kinematic generator for LFD. However, in $(3+1)$ dimensions, the presence of the transverse term $\vec{x}^{\perp}\cdot\vec{x}^{\perp}$ in the numerator of $x^{+\prime}$ (as shown in Table~\ref{tab:Kinematic_and_dynamic_generators_for_different_interoplation_angles_conformal1+1}) means that the light-front time is no longer invariant under $\mathfrak{K}_{-}$. Consequently, $\mathfrak{K}_{-}$ in LFD transforms from being kinematic in $(1+1)$d to becoming dynamical in $(3+1)$d.
However, the longitudinal boost $K_3$
in LFD is still kinematic in (3+1)d.  
As summarized in Table~\ref{tab:Kinematic_and_dynamic_generators_for_different_interoplation_angles_conformal1+1}, out of the 15 conformal generators in $(3+1)$ dimensions, 8 generators are kinematic in LFD, compared to 7 in IFD. This demonstrates that LFD still maximizes the total number of kinematic generators in $(3+1)$ dimensions, albeit with a different dynamical balance than in lower dimensions.

\KinematicGeneratorTable
\FloatBarrier


\section{Summary and Conclusion}
\label{summary-conclusion}

In this work, we presented the interpolating conformal algebra between IFD and LFD in $(3+1)$ dimensions. We elaborated on the $(3+1)$ dimensional conformal algebra using the interpolating projective spacetime representation and showed that LFD maximizes the number of kinematic generators. As discussed in this work, identifying which conformal generators become kinematic in LFD is useful for understanding the utility of LFD in hadron phenomenology, thereby saving considerable dynamical effort when solving QCD.

Building on the correspondence between $(1+1)$d quantum field theory and AdS$_3$, we plan to analyze the holographic dual correspondence between $(3+1)$d QCD and AdS$_5$ by further interpolating between IFD and LFD. With this algebraic framework, we can evaluate light-front QCD, which maximizes the number of kinematic generators and links first-principle QCD to the light-front quark model for hadronic phenomenology. Handling relativistic interactions among hadron constituents is relevant for understanding the advantages of different forms of relativistic dynamics~\cite{Ji2025}.

\acknowledgments{This work was supported by the U.S. Department of Energy (Grant No. DE-FG02-03ER41260).
The National Energy Research Scientific Computing Center (NERSC) supported by the Office of Science of the U.S. Department of Energy under Contract No. DE-AC02-05CH11231 is also acknowledged. 
}
\appendix
\setcounter{section}{0}
\section{Explicit \texorpdfstring{$4\times 4$}{4x4} Matrix Representations}
\label{app:explicit_matrices}
The Dirac matrices obey the Clifford algebra
\begin{align}
\{\gamma^\mu,\gamma^\nu\}=2g^{\mu\nu}\mathbbm{1},\qquad
\gamma_\mu=g_{\mu\nu}\gamma^\nu,\qquad
\gamma_5=i\gamma^0\gamma^1\gamma^2\gamma^3,
\end{align}
where $\mathbbm{1}$ is the $4\times 4$ identity matrix. While this algebra and the resulting isomorphism hold for any general basis, we use the standard Dirac basis as an explicit example. The explicit $4\times 4$ matrix representations are provided below. The following dictionary realizes all fifteen IFD conformal generators
in terms of the corresponding standard Dirac matrices:
\begin{align}
P_\mu&=\frac{i}{2}(\mathbbm{1}+\gamma_5)\gamma_\mu,
&\mathfrak{K}_\mu&=-\frac{i}{2}(\mathbbm{1}-\gamma_5)\gamma_\mu,\nonumber\\
D&=\frac{i}{2}\gamma_5,
&K^i&=\frac{i}{2}\gamma_i\gamma_0,\qquad i=1,2,3,\nonumber\\
J^1&=\frac{i}{2}\gamma_2\gamma_3,
&J^2&=\frac{i}{2}\gamma_3\gamma_1,
\qquad J^3=\frac{i}{2}\gamma_1\gamma_2.
\label{eq:ifd-dirac-dictionary}
\end{align}
The longitudinal generator in the projective array has the lowered-index
form $K_3=-K^3$. Substitution of Eq.~\eqref{eq:ifd-dirac-dictionary} into
Eq.~\eqref{Jab} yields the following representation which we denote as $\rho_\gamma(J_{ab})$: \cite{Dirac1936, Hepner1962, Weinberg_2010}
\begingroup\footnotesize
\begin{align}
\rho_\gamma(J_{ab})=\frac{i}{2}
\begin{pmatrix}
0&-\gamma_5&\dfrac{(\mathbbm{1}-\gamma_5)\gamma_0}{\sqrt{2}}&\dfrac{(\mathbbm{1}-\gamma_5)\gamma_1}{\sqrt{2}}&\dfrac{(\mathbbm{1}-\gamma_5)\gamma_2}{\sqrt{2}}&\dfrac{(\mathbbm{1}-\gamma_5)\gamma_3}{\sqrt{2}}\\
\gamma_5&0&\dfrac{(\mathbbm{1}+\gamma_5)\gamma_0}{\sqrt{2}}&\dfrac{(\mathbbm{1}+\gamma_5)\gamma_1}{\sqrt{2}}&\dfrac{(\mathbbm{1}+\gamma_5)\gamma_2}{\sqrt{2}}&\dfrac{(\mathbbm{1}+\gamma_5)\gamma_3}{\sqrt{2}}\\
-\dfrac{(\mathbbm{1}-\gamma_5)\gamma_0}{\sqrt{2}}&-\dfrac{(\mathbbm{1}+\gamma_5)\gamma_0}{\sqrt{2}}&0&\gamma_0\gamma_1&\gamma_0\gamma_2&\gamma_0\gamma_3\\
-\dfrac{(\mathbbm{1}-\gamma_5)\gamma_1}{\sqrt{2}}&-\dfrac{(\mathbbm{1}+\gamma_5)\gamma_1}{\sqrt{2}}&\gamma_1\gamma_0&0&\gamma_1\gamma_2&\gamma_1\gamma_3\\
-\dfrac{(\mathbbm{1}-\gamma_5)\gamma_2}{\sqrt{2}}&-\dfrac{(\mathbbm{1}+\gamma_5)\gamma_2}{\sqrt{2}}&\gamma_2\gamma_0&\gamma_2\gamma_1&0&\gamma_2\gamma_3\\
-\dfrac{(\mathbbm{1}-\gamma_5)\gamma_3}{\sqrt{2}}&-\dfrac{(\mathbbm{1}+\gamma_5)\gamma_3}{\sqrt{2}}&\gamma_3\gamma_0&\gamma_3\gamma_1&\gamma_3\gamma_2&0
\end{pmatrix}.
\label{eq:ifd-dirac-jab}
\end{align}
\endgroup
For example, $\rho_\gamma(J_{-2,-1})=-i\gamma_5/2=-D$ and the
translation and special-conformal entries reproduce
$J_{-1,\mu}=P_\mu/\sqrt{2}$ and
$J_{-2,\mu}=-\mathfrak{K}_\mu/\sqrt{2}$, respectively. The Clifford
algebra consequently reproduces the IFD commutators in
Eq.~\eqref{algebrasimp} and Table~\ref{tableIFD}.

The standard Dirac basis for the $\gamma$ matrices is given by
\begin{align*}
\gamma^0&=\begin{pmatrix}1&0&0&0\\0&1&0&0\\0&0&-1&0\\0&0&0&-1\end{pmatrix}~;~~~~~~
\gamma^1=\begin{pmatrix}0&0&0&1\\0&0&1&0\\0&-1&0&0\\-1&0&0&0\end{pmatrix};\\
\gamma^2&=\begin{pmatrix}0&0&0&-i\\0&0&i&0\\0&i&0&0\\-i&0&0&0\end{pmatrix}~;~~~~~~
\gamma^3=\begin{pmatrix}0&0&1&0\\0&0&0&-1\\-1&0&0&0\\0&1&0&0\end{pmatrix};\\
\gamma_5&=\begin{pmatrix}0&0&1&0\\0&0&0&1\\1&0&0&0\\0&1&0&0\end{pmatrix}~;~~~~~~
\mathbbm{1}=\begin{pmatrix}1&0&0&0\\0&1&0&0\\0&0&1&0\\0&0&0&1\end{pmatrix}.
\end{align*}

Using the dictionary in Eq.~\eqref{eq:ifd-dirac-dictionary}, the explicit $4\times 4$ representations for the fifteen conformal generators are:
\begin{align*}
    J^{1}&=\frac{1}{2}\begin{pmatrix}
    0&1&0&0\\
    1&0&0&0\\
    0&0&0&1\\
    0&0&1&0
    \end{pmatrix}~;~~~~~~J^{2}=\frac{1}{2}\begin{pmatrix}
    0&-i&0&0\\
    i&0&0&0\\
    0&0&0&-i\\
    0&0&i&0
    \end{pmatrix}~;\\
    J^{3}&=\frac{1}{2}\begin{pmatrix}
    1&0&0&0\\
    0&-1&0&0\\
    0&0&1&0\\
    0&0&0&-1
    \end{pmatrix}~;~~~~~~
    K^{1}=\frac{i}{2}\begin{pmatrix}
    0&0&0&1\\
    0&0&1&0\\
    0&1&0&0\\
    1&0&0&0
    \end{pmatrix}~;\\
    K^{2}&=\frac{1}{2}\begin{pmatrix}
    0&0&0&1\\
    0&0&-1&0\\
    0&1&0&0\\
    -1&0&0&0
    \end{pmatrix}~;~~~~~~K^{3}=\frac{i}{2}\begin{pmatrix}
    0&0&1&0\\
    0&0&0&-1\\
    1&0&0&0\\
    0&-1&0&0
    \end{pmatrix}~;\\
    P_0&=\frac{i}{2}\begin{pmatrix}
    1&0&-1&0\\
    0&1&0&-1\\
    1&0&-1&0\\
    0&1&0&-1
    \end{pmatrix}~;~~~~~~P_1=\frac{i}{2}\begin{pmatrix}
    0&1&0&-1\\
    1&0&-1&0\\
    0&1&0&-1\\
    1&0&-1&0
    \end{pmatrix}~;\\
    P_2&=\frac{1}{2}\begin{pmatrix}
    0&1&0&-1\\
    -1&0&1&0\\
    0&1&0&-1\\
    -1&0&1&0
    \end{pmatrix}~;~~~~~~P_3=\frac{i}{2}\begin{pmatrix}
    1&0&-1&0\\
    0&-1&0&1\\
    1&0&-1&0\\
    0&-1&0&1
    \end{pmatrix}\\
    \mathfrak{K}_0&=\frac{i}{2}\begin{pmatrix}
    -1&0&-1&0\\
    0&-1&0&-1\\
    1&0&1&0\\
    0&1&0&1
    \end{pmatrix}~;~~~~~~\mathfrak{K}_1=\frac{i}{2}\begin{pmatrix}
    0&1&0&1\\
    1&0&1&0\\
    0&-1&0&-1\\
    -1&0&-1&0
    \end{pmatrix}~;\\
    \mathfrak{K}_2&=\frac{1}{2}\begin{pmatrix}
    0&1&0&1\\
    -1&0&-1&0\\
    0&-1&0&-1\\
    1&0&1&0
    \end{pmatrix}~;~~~~~~\mathfrak{K}_3=\frac{i}{2}\begin{pmatrix}
    1&0&1&0\\
    0&-1&0&-1\\
    -1&0&-1&0\\
    0&1&0&1
    \end{pmatrix}\\
    D&=\frac{i}{2}\begin{pmatrix}
    0&0&1&0\\
    0&0&0&1\\
    1&0&0&0\\
    0&1&0&0
    \end{pmatrix}.
\end{align*}

\section{Derivation of the Interpolating-Time Transformations}
\label{app:interpolating-time-derivations}

This appendix gives the essential derivations of the fifteen entries in
Table~\ref{tablexprime}. For each individual transformation, we apply the
corresponding matrix in Eq.~\eqref{matrix-rep-conformal-generators} directly
to the interpolating projective space-time vector. The resulting interpolating
projective space-time components are substituted
in the defining ratios for $x_{\hat{+}}'$ and $x_{\hat{-}}'$, and then
$x^{\hat{+}\prime}$ is obtained by raising the index as in
Section~\ref{sec:projectivespace}.
Only one generator parameter is nonzero in each subsection.

\subsection{Interpolating Longitudinal special conformal transformations (SCTs)}

\subsubsection{\texorpdfstring{$\mathfrak{K}^{\hat{+}}$}{K hat plus}}

The transformation of the interpolating projective space-time vector by
$\mathfrak{K}^{\hat{+}}$ is
\begin{align}
\begin{pmatrix}
X_{-2}'\\ X_{-1}'\\ X_{\hat{+}}'\\ X_1'\\ X_2'\\ X_{\hat{-}}'
\end{pmatrix}
&=e^{-ib^{\hat{+}}\mathfrak{K}^{\hat{+}}}
\begin{pmatrix}
X_{-2}\\ X_{-1}\\ X_{\hat{+}}\\ X_1\\ X_2\\ X_{\hat{-}}
\end{pmatrix}
=\begin{pmatrix}
1 & 0 & 0 & 0 & 0 & 0\\
\mathbb{C} (b^{\hat{+}})^{2} & 1 & \sqrt{2} b^{\hat{+}} & 0 & 0 & 0\\
\sqrt{2} \mathbb{C} b^{\hat{+}} & 0 & 1 & 0 & 0 & 0\\
0 & 0 & 0 & 1 & 0 & 0\\
0 & 0 & 0 & 0 & 1 & 0\\
\sqrt{2} \mathbb{S} b^{\hat{+}} & 0 & 0 & 0 & 0 & 1
\end{pmatrix}
\begin{pmatrix}
X_{-2}\\ X_{-1}\\ X_{\hat{+}}\\ X_1\\ X_2\\ X_{\hat{-}}
\end{pmatrix}\\
&=\begin{pmatrix}
X_{-2}\\
X_{-1}+\sqrt{2}b^{\hat{+}}X_{\hat{+}}
+\mathbb{C}(b^{\hat{+}})^2X_{-2}\\
X_{\hat{+}}+\sqrt{2}\mathbb{C}b^{\hat{+}}X_{-2}\\
X_1\\
X_2\\
X_{\hat{-}}+\sqrt{2}\mathbb{S}b^{\hat{+}}X_{-2}
\end{pmatrix}.
\end{align}
\begin{align}
x_{\hat{+}}'
&=-\frac{1}{\sqrt{2}}\frac{X_{\hat{+}}'}{X_{-1}'}
=-\frac{1}{\sqrt{2}}
\frac{X_{\hat{+}}/X_{-1}+\sqrt{2}\mathbb{C}b^{\hat{+}}X_{-2}/X_{-1}}
{1+\sqrt{2}b^{\hat{+}}X_{\hat{+}}/X_{-1}
 +\mathbb{C}(b^{\hat{+}})^2X_{-2}/X_{-1}}\nonumber\\
&=\frac{x_{\hat{+}}-\mathbb{C}b^{\hat{+}}x^2}
{1-2b^{\hat{+}}x_{\hat{+}}+\mathbb{C}(b^{\hat{+}})^2x^2}, \nonumber\\
x_{\hat{-}}'
&=-\frac{1}{\sqrt{2}}\frac{X_{\hat{-}}'}{X_{-1}'}
=-\frac{1}{\sqrt{2}}
\frac{X_{\hat{-}}/X_{-1}+\sqrt{2}\mathbb{S}b^{\hat{+}}X_{-2}/X_{-1}}
{1+\sqrt{2}b^{\hat{+}}X_{\hat{+}}/X_{-1}
 +\mathbb{C}(b^{\hat{+}})^2X_{-2}/X_{-1}}\nonumber\\
&=\frac{x_{\hat{-}}-\mathbb{S}b^{\hat{+}}x^2}
{1-2b^{\hat{+}}x_{\hat{+}}+\mathbb{C}(b^{\hat{+}})^2x^2}.
\end{align}
Consequently,
\begin{align}
x^{\hat{+}\prime}
&=\mathbb{C}x_{\hat{+}}'+\mathbb{S}x_{\hat{-}}'\nonumber\\
&=\frac{x^{\hat{+}}-b^{\hat{+}}x^2}
{1-2\mathbb{C}b^{\hat{+}}x^{\hat{+}}
 -2\mathbb{S}b^{\hat{+}}x^{\hat{-}}
 +\mathbb{C}(b^{\hat{+}})^2x^2},
\end{align}
which is the first entry of Table~\ref{tablexprime}.

\subsubsection{\texorpdfstring{$\mathfrak{K}^{\hat{-}}$}{K hat minus}}

The transformation of the interpolating projective space-time vector by
$\mathfrak{K}^{\hat{-}}$ is
\begin{align}
\begin{pmatrix}
X_{-2}'\\ X_{-1}'\\ X_{\hat{+}}'\\ X_1'\\ X_2'\\ X_{\hat{-}}'
\end{pmatrix}
&=e^{-ib^{\hat{-}}\mathfrak{K}^{\hat{-}}}
\begin{pmatrix}
X_{-2}\\ X_{-1}\\ X_{\hat{+}}\\ X_1\\ X_2\\ X_{\hat{-}}
\end{pmatrix}
=\begin{pmatrix}
1 & 0 & 0 & 0 & 0 & 0\\
- \mathbb{C} (b^{\hat{-}})^{2} & 1 & 0 & 0 & 0 & \sqrt{2} b^{\hat{-}}\\
\sqrt{2} \mathbb{S} b^{\hat{-}} & 0 & 1 & 0 & 0 & 0\\
0 & 0 & 0 & 1 & 0 & 0\\
0 & 0 & 0 & 0 & 1 & 0\\
- \sqrt{2} \mathbb{C} b^{\hat{-}} & 0 & 0 & 0 & 0 & 1
\end{pmatrix}
\begin{pmatrix}
X_{-2}\\ X_{-1}\\ X_{\hat{+}}\\ X_1\\ X_2\\ X_{\hat{-}}
\end{pmatrix}\\
&=\begin{pmatrix}
X_{-2}\\
X_{-1}+\sqrt{2}b^{\hat{-}}X_{\hat{-}}
-\mathbb{C}(b^{\hat{-}})^2X_{-2}\\
X_{\hat{+}}+\sqrt{2}\mathbb{S}b^{\hat{-}}X_{-2}\\
X_1\\
X_2\\
X_{\hat{-}}-\sqrt{2}\mathbb{C}b^{\hat{-}}X_{-2}
\end{pmatrix}.
\end{align}
\begin{align}
x_{\hat{+}}'
&=-\frac{1}{\sqrt{2}}\frac{X_{\hat{+}}'}{X_{-1}'}
=-\frac{1}{\sqrt{2}}
\frac{X_{\hat{+}}/X_{-1}+\sqrt{2}\mathbb{S}b^{\hat{-}}X_{-2}/X_{-1}}
{1+\sqrt{2}b^{\hat{-}}X_{\hat{-}}/X_{-1}
 -\mathbb{C}(b^{\hat{-}})^2X_{-2}/X_{-1}}\nonumber\\
&=\frac{x_{\hat{+}}-\mathbb{S}b^{\hat{-}}x^2}
{1-2b^{\hat{-}}x_{\hat{-}}-\mathbb{C}(b^{\hat{-}})^2x^2}, \nonumber\\
x_{\hat{-}}'
&=-\frac{1}{\sqrt{2}}\frac{X_{\hat{-}}'}{X_{-1}'}
=-\frac{1}{\sqrt{2}}
\frac{X_{\hat{-}}/X_{-1}-\sqrt{2}\mathbb{C}b^{\hat{-}}X_{-2}/X_{-1}}
{1+\sqrt{2}b^{\hat{-}}X_{\hat{-}}/X_{-1}
 -\mathbb{C}(b^{\hat{-}})^2X_{-2}/X_{-1}}\nonumber\\
&=\frac{x_{\hat{-}}+\mathbb{C}b^{\hat{-}}x^2}
{1-2b^{\hat{-}}x_{\hat{-}}-\mathbb{C}(b^{\hat{-}})^2x^2}.
\end{align}
The $b^{\hat{-}}x^2$ terms cancel when the index is raised:
\begin{align}
x^{\hat{+}\prime}
&=\frac{x^{\hat{+}}}
{1+2\mathbb{C}b^{\hat{-}}x^{\hat{-}}
 -2\mathbb{S}b^{\hat{-}}x^{\hat{+}}
 -\mathbb{C}(b^{\hat{-}})^2x^2}.
\end{align}

\subsection{Interpolating Longitudinal Translations}

\subsubsection{\texorpdfstring{$P_{\hat{+}}$}{P hat plus}}

The transformation of the interpolating projective space-time vector by
$P_{\hat{+}}$ is
\begin{align}
\begin{pmatrix}
X_{-2}'\\ X_{-1}'\\ X_{\hat{+}}'\\ X_1'\\ X_2'\\ X_{\hat{-}}'
\end{pmatrix}
&=e^{-ia^{\hat{+}}P_{\hat{+}}}
\begin{pmatrix}
X_{-2}\\ X_{-1}\\ X_{\hat{+}}\\ X_1\\ X_2\\ X_{\hat{-}}
\end{pmatrix}
=\begin{pmatrix}
1 & \mathbb{C} (a^{\hat{+}})^{2} & - \sqrt{2} a^{\hat{+}} & 0 & 0 & 0\\
0 & 1 & 0 & 0 & 0 & 0\\
0 & - \sqrt{2} \mathbb{C} a^{\hat{+}} & 1 & 0 & 0 & 0\\
0 & 0 & 0 & 1 & 0 & 0\\
0 & 0 & 0 & 0 & 1 & 0\\
0 & - \sqrt{2} \mathbb{S} a^{\hat{+}} & 0 & 0 & 0 & 1
\end{pmatrix}
\begin{pmatrix}
X_{-2}\\ X_{-1}\\ X_{\hat{+}}\\ X_1\\ X_2\\ X_{\hat{-}}
\end{pmatrix}\\
&=\begin{pmatrix}
X_{-2}+\mathbb{C}(a^{\hat{+}})^2X_{-1}
-\sqrt{2}a^{\hat{+}}X_{\hat{+}}\\
X_{-1}\\
X_{\hat{+}}-\sqrt{2}\mathbb{C}a^{\hat{+}}X_{-1}\\
X_1\\ X_2\\
X_{\hat{-}}-\sqrt{2}\mathbb{S}a^{\hat{+}}X_{-1}
\end{pmatrix}.
\end{align}
Therefore,
\begin{align}
x_{\hat{+}}'
&=-\frac{1}{\sqrt{2}}\frac{X_{\hat{+}}-\sqrt{2}\mathbb{C}a^{\hat{+}}X_{-1}}{X_{-1}}
=x_{\hat{+}}+\mathbb{C}a^{\hat{+}}, \nonumber\\
x_{\hat{-}}'
&=-\frac{1}{\sqrt{2}}\frac{X_{\hat{-}}-\sqrt{2}\mathbb{S}a^{\hat{+}}X_{-1}}{X_{-1}}
=x_{\hat{-}}+\mathbb{S}a^{\hat{+}}, \nonumber\\
x^{\hat{+}\prime}
&=\mathbb{C}x_{\hat{+}}'+\mathbb{S}x_{\hat{-}}'
=x^{\hat{+}}+a^{\hat{+}}.
\end{align}

\subsubsection{\texorpdfstring{$P_{\hat{-}}$}{P hat minus}}

The transformation of the interpolating projective space-time vector by
$P_{\hat{-}}$ is
\begin{align}
\begin{pmatrix}
X_{-2}'\\ X_{-1}'\\ X_{\hat{+}}'\\ X_1'\\ X_2'\\ X_{\hat{-}}'
\end{pmatrix}
&=e^{-ia^{\hat{-}}P_{\hat{-}}}
\begin{pmatrix}
X_{-2}\\ X_{-1}\\ X_{\hat{+}}\\ X_1\\ X_2\\ X_{\hat{-}}
\end{pmatrix}
=\begin{pmatrix}
1 & - \mathbb{C} (a^{\hat{-}})^{2} & 0 & 0 & 0 & - \sqrt{2} a^{\hat{-}}\\
0 & 1 & 0 & 0 & 0 & 0\\
0 & - \sqrt{2} \mathbb{S} a^{\hat{-}} & 1 & 0 & 0 & 0\\
0 & 0 & 0 & 1 & 0 & 0\\
0 & 0 & 0 & 0 & 1 & 0\\
0 & \sqrt{2} \mathbb{C} a^{\hat{-}} & 0 & 0 & 0 & 1
\end{pmatrix}
\begin{pmatrix}
X_{-2}\\ X_{-1}\\ X_{\hat{+}}\\ X_1\\ X_2\\ X_{\hat{-}}
\end{pmatrix}\\
&=\begin{pmatrix}
X_{-2}-\mathbb{C}(a^{\hat{-}})^2X_{-1}
-\sqrt{2}a^{\hat{-}}X_{\hat{-}}\\
X_{-1}\\
X_{\hat{+}}-\sqrt{2}\mathbb{S}a^{\hat{-}}X_{-1}\\
X_1\\ X_2\\
X_{\hat{-}}+\sqrt{2}\mathbb{C}a^{\hat{-}}X_{-1}
\end{pmatrix}.
\end{align}
so that
\begin{align}
x_{\hat{+}}'
&=-\frac{1}{\sqrt{2}}\frac{X_{\hat{+}}-\sqrt{2}\mathbb{S}a^{\hat{-}}X_{-1}}{X_{-1}}
=x_{\hat{+}}+\mathbb{S}a^{\hat{-}}, \nonumber\\
x_{\hat{-}}'
&=-\frac{1}{\sqrt{2}}\frac{X_{\hat{-}}+\sqrt{2}\mathbb{C}a^{\hat{-}}X_{-1}}{X_{-1}}
=x_{\hat{-}}-\mathbb{C}a^{\hat{-}}, \nonumber\\
x^{\hat{+}\prime}
&=\mathbb{C}x_{\hat{+}}'+\mathbb{S}x_{\hat{-}}'
=x^{\hat{+}}.
\end{align}

\subsection{Interpolating dilatation transformations}
\subsubsection{\texorpdfstring{$D$}{D}}
The transformation of the interpolating projective
space-time vector by $D$ is
\begin{align}
\begin{pmatrix}
    {X}_{\hat{-2}}'\\
    {X}_{\hat{-1}}'\\
    {X}_{\hat{+}}'\\
    {X}_{\hat{1}}'\\
    {X}_{\hat{2}}'\\
    {X}_{\hat{-}}'\\
    \end{pmatrix}&= \exp{(-i\alpha D)}\begin{pmatrix}
    {X}_{\hat{-2}}\\
    {X}_{\hat{-1}}\\
    {X}_{\hat{+}}\\
    {X}_{\hat{1}}\\
    {X}_{\hat{2}}\\
    {X}_{\hat{-}}\\
    \end{pmatrix}\\
    \begin{pmatrix}
    {X}_{\hat{-2}}'\\
    {X}_{\hat{-1}}'\\
    {X}_{\hat{+}}'\\
    {X}_{\hat{1}}'\\
    {X}_{\hat{2}}'\\
    {X}_{\hat{-}}'\\
    \end{pmatrix}&= \begin{pmatrix}
        e^{-\alpha}&0&0&0&0&0\\
        0&e^{\alpha}&0&0&0&0\\
        0&0&1&0&0&0\\
        0&0&0&1&0&0\\
        0&0&0&0&1&0\\
        0&0&0&0&0&1
    \end{pmatrix}\begin{pmatrix}
    {X}_{\hat{-2}}\\
    {X}_{\hat{-1}}\\
    {X}_{\hat{+}}\\
    {X}_{\hat{1}}\\
    {X}_{\hat{2}}\\
    {X}_{\hat{-}}
    \end{pmatrix}\\
    \begin{pmatrix}
    {X}_{\hat{-2}}'\\
    {X}_{\hat{-1}}'\\
    {X}_{\hat{+}}'\\
    {X}_{\hat{1}}'\\
    {X}_{\hat{2}}'\\
    {X}_{\hat{-}}'
    \end{pmatrix}&= \begin{pmatrix}
    e^{-\alpha}{X}_{-2}\\
    e^{\alpha}{X}_{-1}\\
    {X}_{\hat{+}}\\
    {X}_{\hat{1}}\\
    {X}_{\hat{2}}\\
    {X}_{\hat{-}}
    \end{pmatrix}
\end{align}
On space-time, this transformation gives
\begin{align}
    x_{\hat{+}}'&=\frac{-1}{\sqrt{2}}\frac{{X}_{{\hat{+}}}'}{ {X}_{-1}'}\\
    &=\frac{-1}{\sqrt{2}}\frac{{X}_{{\hat{+}}}}{ e^{\alpha}{X}_{-1}}\\
    x_{\hat{+}}'&=e^{-\alpha}x_{\hat{+}}
\end{align}
and
\begin{align}
    x_{\hat{-}}'&=\frac{-1}{\sqrt{2}}\frac{{X}_{\hat{-}}'}{ {X}_{-1}'}\\
    &=\frac{-1}{\sqrt{2}}\frac{{X}_{\hat{-}}}{ e^{\alpha}{X}_{-1}}\\
    x_{\hat{-}}'&=e^{-\alpha}x_{\hat{-}}
\end{align}
hence
\begin{align}
    x^{\hat{+}\prime}&=e^{-\alpha}x^{\hat{+}}.
\end{align}

\subsubsection{\texorpdfstring{$K_{3}$}{K3}}
The transformation of the interpolating projective
space-time vector by $K_{3}$ is
\begin{align}
\begin{pmatrix}
    {X}_{\hat{-2}}'\\
    {X}_{\hat{-1}}'\\
    {X}_{\hat{+}}'\\
    {X}_{\hat{1}}'\\
    {X}_{\hat{2}}'\\
    {X}_{\hat{-}}'\\
    \end{pmatrix}&= \exp{(-i\eta_{3}K_{3})}\begin{pmatrix}
    {X}_{\hat{-2}}\\
    {X}_{\hat{-1}}\\
    {X}_{\hat{+}}\\
    {X}_{\hat{1}}\\
    {X}_{\hat{2}}\\
    {X}_{\hat{-}}\\
    \end{pmatrix}\\
    \begin{pmatrix}
    {X}_{\hat{-2}}'\\
    {X}_{\hat{-1}}'\\
    {X}_{\hat{+}}'\\
    {X}_{\hat{1}}'\\
    {X}_{\hat{2}}'\\
    {X}_{\hat{-}}'\\
    \end{pmatrix}&= \begin{pmatrix}
        1&0&0&0&0&0\\
        0&1&0&0&0&0\\
        0&0&\cosh{\eta_{3}}+\mathbb{S}\sinh{\eta_{3}}&0&0&-\mathbb{C}\sinh{\eta_{3}}\\
        0&0&0&1&0&0\\
        0&0&0&0&1&0\\
        0&0&-\mathbb{C}\sinh{\eta_{3}}&0&0&\cosh{\eta_{3}}-\mathbb{S}\sinh{\eta_{3}}
    \end{pmatrix}\begin{pmatrix}
    {X}_{\hat{-2}}\\
    {X}_{\hat{-1}}\\
    {X}_{\hat{+}}\\
    {X}_{\hat{1}}\\
    {X}_{\hat{2}}\\
    {X}_{\hat{-}}
    \end{pmatrix}\\
    \begin{pmatrix}
    {X}_{\hat{-2}}'\\
    {X}_{\hat{-1}}'\\
    {X}_{\hat{+}}'\\
    {X}_{\hat{1}}'\\
    {X}_{\hat{2}}'\\
    {X}_{\hat{-}}'
    \end{pmatrix}&= \begin{pmatrix}
        {X}_{\hat{-2}}\\
        {X}_{\hat{-1}}\\
        (\cosh{\eta_{3}}+\mathbb{S}\sinh{\eta_{3}}){X}_{\hat{+}}-(\mathbb{C}\sinh{\eta_{3}}){X}_{\hat{-}}\\
    {X}_{\hat{1}}\\
    {X}_{\hat{2}}\\
        -(\mathbb{C}\sinh{\eta_{3}}){X}_{\hat{+}}+(\cosh{\eta_{3}}-\mathbb{S}\sinh{\eta_{3}}){X}_{\hat{-}}
    \end{pmatrix}
\end{align}
On space-time, this transformation gives
\begin{align}
    x_{\hat{+}}'&=\frac{-1}{\sqrt{2}}\frac{{X}_{{\hat{+}}}'}{ {X}_{-1}'}\\
    &=\frac{-1}{\sqrt{2}}\frac{((\cosh{\eta_{3}}+\mathbb{S}\sinh{\eta_{3}}){X}_{\hat{+}}-(\mathbb{C}\sinh{\eta_{3}}){X}_{\hat{-}})}{{X}_{-1}}\\
    x_{\hat{+}}'&=(\cosh{\eta_{3}}+\mathbb{S}\sinh{\eta_{3}}){x}_{\hat{+}}-(\mathbb{C}\sinh{\eta_{3}}){x}_{\hat{-}}
\end{align}
and
\begin{align}
    x_{\hat{-}}'&=\frac{-1}{\sqrt{2}}\frac{{X}_{{\hat{-}}}'}{ {X}_{-1}'}\\
    &=\frac{-1}{\sqrt{2}}\frac{(-(\mathbb{C}\sinh{\eta_{3}}){X}_{\hat{+}}+(\cosh{\eta_{3}}-\mathbb{S}\sinh{\eta_{3}}){X}_{\hat{-}})}{{X}_{-1}}\\
    x_{\hat{-}}'&=-(\mathbb{C}\sinh{\eta_{3}}){x}_{\hat{+}}+(\cosh{\eta_{3}}-\mathbb{S}\sinh{\eta_{3}}){x}_{\hat{-}}
\end{align}
hence
\begin{align}
    x^{\hat{+}\prime}=&\mathbb{C}x^{\prime}_{\hat{+}}+\mathbb{S}x^{\prime}_{\hat{-}}\\
    =&\cosh{\eta_{3}}\mathbb{C}{x}_{\hat{+}}+\sinh{\eta_{3}}\mathbb{S}\mathbb{C}{x}_{\hat{+}}-\sinh{\eta_{3}}\mathbb{C}\mathbb{C}{x}_{\hat{-}}\nonumber\\
    &~~~~~~~~~~~~~~-\sinh{\eta_{3}}\mathbb{S}\mathbb{C}{x}_{\hat{+}}+\cosh{\eta_{3}}\mathbb{S}{x}_{\hat{-}}-\sinh{\eta_{3}}\mathbb{S}\mathbb{S}{x}_{\hat{-}}\\
    =&\cosh{\eta_{3}}(\mathbb{C}x_{\hat{+}}+\mathbb{S}x_{\hat{-}})+\sinh{\eta_{3}}\mathbb{C}(-\mathbb{C}{x}_{\hat{-}}+\mathbb{S}{x}_{\hat{+}})\nonumber\\
    &~~~~~~~~~~~~~~-\sinh{\eta_{3}}\mathbb{S}(\mathbb{C}x_{\hat{+}}+\mathbb{S}x_{\hat{-}})\\
    x^{\hat{+}\prime}=&(\cosh{\eta_{3}}-\mathbb{S}\sinh{\eta_{3}})x^{\hat{+}}+\mathbb{C}(\sinh{\eta_{3}})x^{\hat{-}}
\end{align}

\subsubsection{\texorpdfstring{$D_{\hat{+}}$}{D hat minus}}

Let $v=\alpha_{\hat{+}}\cos\delta$. The transformation of the interpolating projective
space-time vector by $D_{\hat{+}}$ is
\begin{align}
\begin{pmatrix}
X_{-2}'\\ X_{-1}'\\ X_{\hat{+}}'\\ X_1'\\ X_2'\\ X_{\hat{-}}'
\end{pmatrix}
&=e^{-i\alpha_{\hat{+}}D_{\hat{+}}}
\begin{pmatrix}
X_{-2}\\ X_{-1}\\ X_{\hat{+}}\\ X_1\\ X_2\\ X_{\hat{-}}
\end{pmatrix}\\
&=\begin{pmatrix}
e^{-\alpha_{\hat{+}}\sin\delta} & 0 & 0 & 0 & 0 & 0\\
0 & e^{\alpha_{\hat{+}}\sin\delta} & 0 & 0 & 0 & 0\\
0 & 0 & \cosh v+\mathbb{S}\sinh v & 0 & 0 & -\mathbb{C}\sinh v\\
0 & 0 & 0 & 1 & 0 & 0\\
0 & 0 & 0 & 0 & 1 & 0\\
0 & 0 & -\mathbb{C}\sinh v & 0 & 0 & \cosh v-\mathbb{S}\sinh v
\end{pmatrix}
\begin{pmatrix}
X_{-2}\\ X_{-1}\\ X_{\hat{+}}\\ X_1\\ X_2\\ X_{\hat{-}}
\end{pmatrix}\\
&=\begin{pmatrix}
e^{-\alpha_{\hat{+}}\sin\delta}X_{-2}\\
e^{\alpha_{\hat{+}}\sin\delta}X_{-1}\\
\left(\cosh v+\mathbb{S}\sinh v\right)X_{\hat{+}}
 -\mathbb{C}\sinh v\,X_{\hat{-}}\\
X_1\\ X_2\\
-\mathbb{C}\sinh v\,X_{\hat{+}}
 +\left(\cosh v-\mathbb{S}\sinh v\right)X_{\hat{-}}
\end{pmatrix}.
\end{align}
Thus,
\begin{align}
x_{\hat{+}}'&=e^{-\alpha_{\hat{+}}\sin\delta}
\left[\left(\cosh v+\mathbb{S}\sinh v\right)x_{\hat{+}}
 -\mathbb{C}\sinh v\,x_{\hat{-}}\right], \nonumber\\
x_{\hat{-}}'&=e^{-\alpha_{\hat{+}}\sin\delta}
\left[-\mathbb{C}\sinh v\,x_{\hat{+}}
 +\left(\cosh v-\mathbb{S}\sinh v\right)x_{\hat{-}}\right],
\end{align}
and hence
\begin{align}
x^{\hat{+}\prime}
=e^{-\alpha_{\hat{+}}\sin\delta}
\left[\left(\cosh v-\mathbb{S}\sinh v\right)x^{\hat{+}}
 +\mathbb{C}\sinh v\,x^{\hat{-}}\right].
\end{align}

\subsubsection{\texorpdfstring{$D_{\hat{-}}$}{D hat plus}}

Let $u=\alpha_{\hat{-}}\sin\delta$. The transformation of the interpolating projective
space-time vector by $D_{\hat{-}}$ is
\begin{align}
\begin{pmatrix}
X_{-2}'\\ X_{-1}'\\ X_{\hat{+}}'\\ X_1'\\ X_2'\\ X_{\hat{-}}'
\end{pmatrix}
&=e^{-i\alpha_{\hat{-}}D_{\hat{-}}}
\begin{pmatrix}
X_{-2}\\ X_{-1}\\ X_{\hat{+}}\\ X_1\\ X_2\\ X_{\hat{-}}
\end{pmatrix}\\
&=\begin{pmatrix}
e^{-\alpha_{\hat{-}}\cos\delta} & 0 & 0 & 0 & 0 & 0\\
0 & e^{\alpha_{\hat{-}}\cos\delta} & 0 & 0 & 0 & 0\\
0 & 0 & \cosh u-\mathbb{S}\sinh u & 0 & 0 & \mathbb{C}\sinh u\\
0 & 0 & 0 & 1 & 0 & 0\\
0 & 0 & 0 & 0 & 1 & 0\\
0 & 0 & \mathbb{C}\sinh u & 0 & 0 & \cosh u+\mathbb{S}\sinh u
\end{pmatrix}
\begin{pmatrix}
X_{-2}\\ X_{-1}\\ X_{\hat{+}}\\ X_1\\ X_2\\ X_{\hat{-}}
\end{pmatrix}\\
&=\begin{pmatrix}
e^{-\alpha_{\hat{-}}\cos\delta}X_{-2}\\
e^{\alpha_{\hat{-}}\cos\delta}X_{-1}\\
\left(\cosh u-\mathbb{S}\sinh u\right)X_{\hat{+}}
 +\mathbb{C}\sinh u\,X_{\hat{-}}\\
X_1\\ X_2\\
\mathbb{C}\sinh u\,X_{\hat{+}}
 +\left(\cosh u+\mathbb{S}\sinh u\right)X_{\hat{-}}
\end{pmatrix}.
\end{align}
It follows that
\begin{align}
x_{\hat{+}}'&=e^{-\alpha_{\hat{-}}\cos\delta}
\left[\left(\cosh u-\mathbb{S}\sinh u\right)x_{\hat{+}}
 +\mathbb{C}\sinh u\,x_{\hat{-}}\right], \nonumber\\
x_{\hat{-}}'&=e^{-\alpha_{\hat{-}}\cos\delta}
\left[\mathbb{C}\sinh u\,x_{\hat{+}}
 +\left(\cosh u+\mathbb{S}\sinh u\right)x_{\hat{-}}\right].
\end{align}
Raising the index yields
\begin{align}
x^{\hat{+}\prime}
=e^{-\alpha_{\hat{-}}\cos\delta}
\left[\left(\cosh u+\mathbb{S}\sinh u\right)x^{\hat{+}}
 -\mathbb{C}\sinh u\,x^{\hat{-}}\right].
\end{align}

\subsection{Rotation, Transverse Special Conformal Transformations (SCTs), and Transverse translations}

\subsubsection{\texorpdfstring{$J^{3}$}{J3}}

The transformation of the interpolating projective space-time vector by
$J^3$ is
\begin{align}
\begin{pmatrix}
X_{-2}'\\ X_{-1}'\\ X_{\hat{+}}'\\ X_1'\\ X_2'\\ X_{\hat{-}}'
\end{pmatrix}
&=e^{-i\theta J^3}
\begin{pmatrix}
X_{-2}\\ X_{-1}\\ X_{\hat{+}}\\ X_1\\ X_2\\ X_{\hat{-}}
\end{pmatrix}
=\begin{pmatrix}
1 & 0 & 0 & 0 & 0 & 0\\
0 & 1 & 0 & 0 & 0 & 0\\
0 & 0 & 1 & 0 & 0 & 0\\
0 & 0 & 0 & \cos\theta & -\sin\theta & 0\\
0 & 0 & 0 & \sin\theta & \cos\theta & 0\\
0 & 0 & 0 & 0 & 0 & 1
\end{pmatrix}
\begin{pmatrix}
X_{-2}\\ X_{-1}\\ X_{\hat{+}}\\ X_1\\ X_2\\ X_{\hat{-}}
\end{pmatrix}\\
&=\begin{pmatrix}
X_{-2}\\
X_{-1}\\
X_{\hat{+}}\\
\cos\theta X_1-\sin\theta X_2\\
\sin\theta X_1+\cos\theta X_2\\
X_{\hat{-}}
\end{pmatrix}.
\end{align}
and
\begin{align}
x_{\hat{+}}'=-\frac{1}{\sqrt{2}}\frac{X_{\hat{+}}}{X_{-1}}=x_{\hat{+}},\qquad
x_{\hat{-}}'=-\frac{1}{\sqrt{2}}\frac{X_{\hat{-}}}{X_{-1}}=x_{\hat{-}},\qquad
x^{\hat{+}\prime}=x^{\hat{+}}.
\end{align}

\subsubsection{\texorpdfstring{$\mathfrak{K}_{1}$}{K1}}

The transformation of the interpolating projective space-time vector by
$\mathfrak{K}_1$ is
\begin{align}
\begin{pmatrix}
X_{-2}'\\ X_{-1}'\\ X_{\hat{+}}'\\ X_1'\\ X_2'\\ X_{\hat{-}}'
\end{pmatrix}
&=e^{-ib^1\mathfrak{K}_1}
\begin{pmatrix}
X_{-2}\\ X_{-1}\\ X_{\hat{+}}\\ X_1\\ X_2\\ X_{\hat{-}}
\end{pmatrix}
=\begin{pmatrix}
1 & 0 & 0 & 0 & 0 & 0\\
- (b^{1})^{2} & 1 & 0 & \sqrt{2} b^{1} & 0 & 0\\
0 & 0 & 1 & 0 & 0 & 0\\
- \sqrt{2} b^{1} & 0 & 0 & 1 & 0 & 0\\
0 & 0 & 0 & 0 & 1 & 0\\
0 & 0 & 0 & 0 & 0 & 1
\end{pmatrix}
\begin{pmatrix}
X_{-2}\\ X_{-1}\\ X_{\hat{+}}\\ X_1\\ X_2\\ X_{\hat{-}}
\end{pmatrix}\\
&=\begin{pmatrix}
X_{-2}\\
X_{-1}+\sqrt{2}b^1X_1-(b^1)^2X_{-2}\\
X_{\hat{+}}\\
X_1-\sqrt{2}b^1X_{-2}\\
X_2\\
X_{\hat{-}}
\end{pmatrix}.
\end{align}
\begin{align}
x_{\hat{+}}'
&=-\frac{1}{\sqrt{2}}\frac{X_{\hat{+}}'}{X_{-1}'}
=-\frac{1}{\sqrt{2}}
\frac{X_{\hat{+}}/X_{-1}}
{1+\sqrt{2}b^1X_1/X_{-1}-(b^1)^2X_{-2}/X_{-1}}\nonumber\\
&=\frac{x_{\hat{+}}}{1-2b^1x_1-(b^1)^2x^2}, \nonumber\\
x_{\hat{-}}'
&=-\frac{1}{\sqrt{2}}\frac{X_{\hat{-}}'}{X_{-1}'}
=-\frac{1}{\sqrt{2}}
\frac{X_{\hat{-}}/X_{-1}}
{1+\sqrt{2}b^1X_1/X_{-1}-(b^1)^2X_{-2}/X_{-1}}\nonumber\\
&=\frac{x_{\hat{-}}}{1-2b^1x_1-(b^1)^2x^2}, \nonumber\\
x^{\hat{+}\prime}
&=\mathbb{C}x_{\hat{+}}'+\mathbb{S}x_{\hat{-}}'
=\frac{x^{\hat{+}}}{1-2b^1x_1-(b^1)^2x^2}.
\end{align}

\subsubsection{\texorpdfstring{$\mathfrak{K}_{2}$}{K2}}

The transformation of the interpolating projective space-time vector by
$\mathfrak{K}_2$ is
\begin{align}
\begin{pmatrix}
X_{-2}'\\ X_{-1}'\\ X_{\hat{+}}'\\ X_1'\\ X_2'\\ X_{\hat{-}}'
\end{pmatrix}
&=e^{-ib^2\mathfrak{K}_2}
\begin{pmatrix}
X_{-2}\\ X_{-1}\\ X_{\hat{+}}\\ X_1\\ X_2\\ X_{\hat{-}}
\end{pmatrix}
=\begin{pmatrix}
1 & 0 & 0 & 0 & 0 & 0\\
- (b^{2})^{2} & 1 & 0 & 0 & \sqrt{2} b^{2} & 0\\
0 & 0 & 1 & 0 & 0 & 0\\
0 & 0 & 0 & 1 & 0 & 0\\
- \sqrt{2} b^{2} & 0 & 0 & 0 & 1 & 0\\
0 & 0 & 0 & 0 & 0 & 1
\end{pmatrix}
\begin{pmatrix}
X_{-2}\\ X_{-1}\\ X_{\hat{+}}\\ X_1\\ X_2\\ X_{\hat{-}}
\end{pmatrix}\\
&=\begin{pmatrix}
X_{-2}\\
X_{-1}+\sqrt{2}b^2X_2-(b^2)^2X_{-2}\\
X_{\hat{+}}\\
X_1\\
X_2-\sqrt{2}b^2X_{-2}\\
X_{\hat{-}}
\end{pmatrix}.
\end{align}
\begin{align}
x_{\hat{+}}'
&=-\frac{1}{\sqrt{2}}\frac{X_{\hat{+}}'}{X_{-1}'}
=-\frac{1}{\sqrt{2}}
\frac{X_{\hat{+}}/X_{-1}}
{1+\sqrt{2}b^2X_2/X_{-1}-(b^2)^2X_{-2}/X_{-1}}\nonumber\\
&=\frac{x_{\hat{+}}}{1-2b^2x_2-(b^2)^2x^2}, \nonumber\\
x_{\hat{-}}'
&=-\frac{1}{\sqrt{2}}\frac{X_{\hat{-}}'}{X_{-1}'}
=-\frac{1}{\sqrt{2}}
\frac{X_{\hat{-}}/X_{-1}}
{1+\sqrt{2}b^2X_2/X_{-1}-(b^2)^2X_{-2}/X_{-1}}\nonumber\\
&=\frac{x_{\hat{-}}}{1-2b^2x_2-(b^2)^2x^2}, \nonumber\\
x^{\hat{+}\prime}
&=\mathbb{C}x_{\hat{+}}'+\mathbb{S}x_{\hat{-}}'
=\frac{x^{\hat{+}}}{1-2b^2x_2-(b^2)^2x^2}.
\end{align}

\subsubsection{\texorpdfstring{$P_{1}$}{P1}}

The transformation of the interpolating projective space-time vector by
$P_1$ is
\begin{align}
\begin{pmatrix}
X_{-2}'\\ X_{-1}'\\ X_{\hat{+}}'\\ X_1'\\ X_2'\\ X_{\hat{-}}'
\end{pmatrix}
&=e^{-ia^1P_1}
\begin{pmatrix}
X_{-2}\\ X_{-1}\\ X_{\hat{+}}\\ X_1\\ X_2\\ X_{\hat{-}}
\end{pmatrix}
=\begin{pmatrix}
1 & - (a^{1})^{2} & 0 & - \sqrt{2} a^{1} & 0 & 0\\
0 & 1 & 0 & 0 & 0 & 0\\
0 & 0 & 1 & 0 & 0 & 0\\
0 & \sqrt{2} a^{1} & 0 & 1 & 0 & 0\\
0 & 0 & 0 & 0 & 1 & 0\\
0 & 0 & 0 & 0 & 0 & 1
\end{pmatrix}
\begin{pmatrix}
X_{-2}\\ X_{-1}\\ X_{\hat{+}}\\ X_1\\ X_2\\ X_{\hat{-}}
\end{pmatrix}\\
&=\begin{pmatrix}
X_{-2}-(a^1)^2X_{-1}-\sqrt{2}a^1X_1\\
X_{-1}\\ X_{\hat{+}}\\ X_1+\sqrt{2}a^1X_{-1}\\ X_2\\ X_{\hat{-}}
\end{pmatrix}.
\end{align}
\begin{align}
x_{\hat{+}}'=-\frac{1}{\sqrt{2}}\frac{X_{\hat{+}}}{X_{-1}}=x_{\hat{+}},\qquad
x_{\hat{-}}'=-\frac{1}{\sqrt{2}}\frac{X_{\hat{-}}}{X_{-1}}=x_{\hat{-}},\qquad
x^{\hat{+}\prime}=x^{\hat{+}}.
\end{align}

\subsubsection{\texorpdfstring{$P_{2}$}{P2}}

The transformation of the interpolating projective space-time vector by
$P_2$ is
\begin{align}
\begin{pmatrix}
X_{-2}'\\ X_{-1}'\\ X_{\hat{+}}'\\ X_1'\\ X_2'\\ X_{\hat{-}}'
\end{pmatrix}
&=e^{-ia^2P_2}
\begin{pmatrix}
X_{-2}\\ X_{-1}\\ X_{\hat{+}}\\ X_1\\ X_2\\ X_{\hat{-}}
\end{pmatrix}
=\begin{pmatrix}
1 & - (a^{2})^{2} & 0 & 0 & - \sqrt{2} a^{2} & 0\\
0 & 1 & 0 & 0 & 0 & 0\\
0 & 0 & 1 & 0 & 0 & 0\\
0 & 0 & 0 & 1 & 0 & 0\\
0 & \sqrt{2} a^{2} & 0 & 0 & 1 & 0\\
0 & 0 & 0 & 0 & 0 & 1
\end{pmatrix}
\begin{pmatrix}
X_{-2}\\ X_{-1}\\ X_{\hat{+}}\\ X_1\\ X_2\\ X_{\hat{-}}
\end{pmatrix}\\
&=\begin{pmatrix}
X_{-2}-(a^2)^2X_{-1}-\sqrt{2}a^2X_2\\
X_{-1}\\ X_{\hat{+}}\\ X_1\\ X_2+\sqrt{2}a^2X_{-1}\\ X_{\hat{-}}
\end{pmatrix}.
\end{align}
\begin{align}
x_{\hat{+}}'=-\frac{1}{\sqrt{2}}\frac{X_{\hat{+}}}{X_{-1}}=x_{\hat{+}},\qquad
x_{\hat{-}}'=-\frac{1}{\sqrt{2}}\frac{X_{\hat{-}}}{X_{-1}}=x_{\hat{-}},\qquad
x^{\hat{+}\prime}=x^{\hat{+}}.
\end{align}

\subsection{Transverse interpolating boosts}

\subsubsection{\texorpdfstring{$\mathcal{D}^{\hat{1}}$}{D1}}

Set $q_1=\sqrt{\mathbb{C}}\eta^{1}$. The transformation of the interpolating
projective space-time vector by $\mathcal{D}^{\hat{1}}$ is
\begin{align}
\begin{pmatrix}
X_{-2}'\\ X_{-1}'\\ X_{\hat{+}}'\\ X_1'\\ X_2'\\ X_{\hat{-}}'
\end{pmatrix}
&=e^{-i\eta^{1}\mathcal{D}^{\hat{1}}}
\begin{pmatrix}
X_{-2}\\ X_{-1}\\ X_{\hat{+}}\\ X_1\\ X_2\\ X_{\hat{-}}
\end{pmatrix}
=\begin{pmatrix}
1 & 0 & 0 & 0 & 0 & 0\\
0 & 1 & 0 & 0 & 0 & 0\\
0 & 0 & \cosh q_1 & \sqrt{\mathbb{C}}\sinh q_1 & 0 & 0\\
0 & 0 & \dfrac{\sinh q_1}{\sqrt{\mathbb{C}}} & \cosh q_1 & 0 & 0\\
0 & 0 & 0 & 0 & 1 & 0\\
0 & 0 & \dfrac{\mathbb{S}(\cosh q_1-1)}{\mathbb{C}} & \dfrac{\mathbb{S}\sinh q_1}{\sqrt{\mathbb{C}}} & 0 & 1
\end{pmatrix}
\begin{pmatrix}
X_{-2}\\ X_{-1}\\ X_{\hat{+}}\\ X_1\\ X_2\\ X_{\hat{-}}
\end{pmatrix}\\
&=\begin{pmatrix}
X_{-2}\\ X_{-1}\\
\cosh q_1\,X_{\hat{+}}+\sqrt{\mathbb{C}}\sinh q_1\,X_1\\
\dfrac{\sinh q_1}{\sqrt{\mathbb{C}}}X_{\hat{+}}+\cosh q_1\,X_1\\
X_2\\
X_{\hat{-}}+\dfrac{\mathbb{S}(\cosh q_1-1)}{\mathbb{C}}X_{\hat{+}}
+\dfrac{\mathbb{S}\sinh q_1}{\sqrt{\mathbb{C}}}X_1
\end{pmatrix}.
\end{align}
Therefore,
\begin{align}
x_{\hat{+}}'&=\cosh q_1\,x_{\hat{+}}
 +\sqrt{\mathbb{C}}\sinh q_1\,x_1, \nonumber\\
x_{\hat{-}}'&=x_{\hat{-}}
 +\frac{\mathbb{S}(\cosh q_1-1)}{\mathbb{C}}x_{\hat{+}}
 +\frac{\mathbb{S}\sinh q_1}{\sqrt{\mathbb{C}}}x_1.
\end{align}
Using the interpolating metric to raise the index gives
\begin{align}
x^{\hat{+}\prime}
&=\cosh q_1\,x^{\hat{+}}
 +\frac{\mathbb{S}}{\mathbb{C}}\left(\cosh q_1-1\right)x^{\hat{-}}
 +\frac{\sinh q_1}{\sqrt{\mathbb{C}}}x_1.
\end{align}

\subsubsection{\texorpdfstring{$\mathcal{D}^{\hat{2}}$}{D2}}

With $q_2=\sqrt{\mathbb{C}}\eta^{2}$, the transformation of the interpolating
projective space-time vector by $\mathcal{D}^{\hat{2}}$ is
\begin{align}
\begin{pmatrix}
X_{-2}'\\ X_{-1}'\\ X_{\hat{+}}'\\ X_1'\\ X_2'\\ X_{\hat{-}}'
\end{pmatrix}
&=e^{-i\eta^{2}\mathcal{D}^{\hat{2}}}
\begin{pmatrix}
X_{-2}\\ X_{-1}\\ X_{\hat{+}}\\ X_1\\ X_2\\ X_{\hat{-}}
\end{pmatrix}
=\begin{pmatrix}
1 & 0 & 0 & 0 & 0 & 0\\
0 & 1 & 0 & 0 & 0 & 0\\
0 & 0 & \cosh q_2 & 0 & \sqrt{\mathbb{C}}\sinh q_2 & 0\\
0 & 0 & 0 & 1 & 0 & 0\\
0 & 0 & \dfrac{\sinh q_2}{\sqrt{\mathbb{C}}} & 0 & \cosh q_2 & 0\\
0 & 0 & \dfrac{\mathbb{S}(\cosh q_2-1)}{\mathbb{C}} & 0 & \dfrac{\mathbb{S}\sinh q_2}{\sqrt{\mathbb{C}}} & 1
\end{pmatrix}
\begin{pmatrix}
X_{-2}\\ X_{-1}\\ X_{\hat{+}}\\ X_1\\ X_2\\ X_{\hat{-}}
\end{pmatrix}\\
&=\begin{pmatrix}
X_{-2}\\ X_{-1}\\
\cosh q_2\,X_{\hat{+}}+\sqrt{\mathbb{C}}\sinh q_2\,X_2\\
X_1\\
\dfrac{\sinh q_2}{\sqrt{\mathbb{C}}}X_{\hat{+}}+\cosh q_2\,X_2\\
X_{\hat{-}}+\dfrac{\mathbb{S}(\cosh q_2-1)}{\mathbb{C}}X_{\hat{+}}
+\dfrac{\mathbb{S}\sinh q_2}{\sqrt{\mathbb{C}}}X_2
\end{pmatrix}.
\end{align}
Hence,
\begin{align}
x_{\hat{+}}'&=\cosh q_2\,x_{\hat{+}}
 +\sqrt{\mathbb{C}}\sinh q_2\,x_2, \nonumber\\
x_{\hat{-}}'&=x_{\hat{-}}
 +\frac{\mathbb{S}(\cosh q_2-1)}{\mathbb{C}}x_{\hat{+}}
 +\frac{\mathbb{S}\sinh q_2}{\sqrt{\mathbb{C}}}x_2,
\end{align}
and
\begin{align}
x^{\hat{+}\prime}
&=\cosh q_2\,x^{\hat{+}}
 +\frac{\mathbb{S}}{\mathbb{C}}\left(\cosh q_2-1\right)x^{\hat{-}}
 +\frac{\sinh q_2}{\sqrt{\mathbb{C}}}x_2.
\end{align}

\subsection{Transverse interpolating rotations}

\subsubsection{\texorpdfstring{$\mathcal{K}^{\hat{1}}$}{K1}}

Let $r_1=\sqrt{\mathbb{C}}\phi^{1}$. The transformation of the interpolating
projective space-time vector by $\mathcal{K}^{\hat{1}}$ is
\begin{align}
\begin{pmatrix}
X_{-2}'\\ X_{-1}'\\ X_{\hat{+}}'\\ X_1'\\ X_2'\\ X_{\hat{-}}'
\end{pmatrix}
&=e^{-i\phi^{1}\mathcal{K}^{\hat{1}}}
\begin{pmatrix}
X_{-2}\\ X_{-1}\\ X_{\hat{+}}\\ X_1\\ X_2\\ X_{\hat{-}}
\end{pmatrix}
=\begin{pmatrix}
1 & 0 & 0 & 0 & 0 & 0\\
0 & 1 & 0 & 0 & 0 & 0\\
0 & 0 & 1 & \dfrac{\mathbb{S}\sin r_1}{\sqrt{\mathbb{C}}} & 0 & \dfrac{\mathbb{S}(1-\cos r_1)}{\mathbb{C}}\\
0 & 0 & 0 & \cos r_1 & 0 & \dfrac{\sin r_1}{\sqrt{\mathbb{C}}}\\
0 & 0 & 0 & 0 & 1 & 0\\
0 & 0 & 0 & -\sqrt{\mathbb{C}}\sin r_1 & 0 & \cos r_1
\end{pmatrix}
\begin{pmatrix}
X_{-2}\\ X_{-1}\\ X_{\hat{+}}\\ X_1\\ X_2\\ X_{\hat{-}}
\end{pmatrix}\\
&=\begin{pmatrix}
X_{-2}\\ X_{-1}\\
X_{\hat{+}}+\dfrac{\mathbb{S}\sin r_1}{\sqrt{\mathbb{C}}}X_1
+\dfrac{\mathbb{S}(1-\cos r_1)}{\mathbb{C}}X_{\hat{-}}\\
\cos r_1\,X_1+\dfrac{\sin r_1}{\sqrt{\mathbb{C}}}X_{\hat{-}}\\
X_2\\
-\sqrt{\mathbb{C}}\sin r_1\,X_1+\cos r_1\,X_{\hat{-}}
\end{pmatrix}.
\end{align}
The transformed lower coordinates are
\begin{align}
x_{\hat{+}}'&=x_{\hat{+}}
 +\frac{\mathbb{S}\sin r_1}{\sqrt{\mathbb{C}}}x_1
 +\frac{\mathbb{S}(1-\cos r_1)}{\mathbb{C}}x_{\hat{-}}, \nonumber\\
x_{\hat{-}}'&=-\sqrt{\mathbb{C}}\sin r_1\,x_1
 +\cos r_1\,x_{\hat{-}}.
\end{align}
The transverse and $x_{\hat{-}}$ contributions cancel after raising the
index:
\begin{align}
x^{\hat{+}\prime}
&=\mathbb{C}x_{\hat{+}}'+\mathbb{S}x_{\hat{-}}'
=\mathbb{C}x_{\hat{+}}+\mathbb{S}x_{\hat{-}}
=x^{\hat{+}}.
\end{align}

\subsubsection{\texorpdfstring{$\mathcal{K}^{\hat{2}}$}{K2}}

For $r_2=\sqrt{\mathbb{C}}\phi^{2}$, the transformation of the interpolating
projective space-time vector by $\mathcal{K}^{\hat{2}}$ is
\begin{align}
\begin{pmatrix}
X_{-2}'\\ X_{-1}'\\ X_{\hat{+}}'\\ X_1'\\ X_2'\\ X_{\hat{-}}'
\end{pmatrix}
&=e^{-i\phi^{2}\mathcal{K}^{\hat{2}}}
\begin{pmatrix}
X_{-2}\\ X_{-1}\\ X_{\hat{+}}\\ X_1\\ X_2\\ X_{\hat{-}}
\end{pmatrix}
=\begin{pmatrix}
1 & 0 & 0 & 0 & 0 & 0\\
0 & 1 & 0 & 0 & 0 & 0\\
0 & 0 & 1 & 0 & \dfrac{\mathbb{S}\sin r_2}{\sqrt{\mathbb{C}}} & \dfrac{\mathbb{S}(1-\cos r_2)}{\mathbb{C}}\\
0 & 0 & 0 & 1 & 0 & 0\\
0 & 0 & 0 & 0 & \cos r_2 & \dfrac{\sin r_2}{\sqrt{\mathbb{C}}}\\
0 & 0 & 0 & 0 & -\sqrt{\mathbb{C}}\sin r_2 & \cos r_2
\end{pmatrix}
\begin{pmatrix}
X_{-2}\\ X_{-1}\\ X_{\hat{+}}\\ X_1\\ X_2\\ X_{\hat{-}}
\end{pmatrix}\\
&=\begin{pmatrix}
X_{-2}\\ X_{-1}\\
X_{\hat{+}}+\dfrac{\mathbb{S}\sin r_2}{\sqrt{\mathbb{C}}}X_2
+\dfrac{\mathbb{S}(1-\cos r_2)}{\mathbb{C}}X_{\hat{-}}\\
X_1\\
\cos r_2\,X_2+\dfrac{\sin r_2}{\sqrt{\mathbb{C}}}X_{\hat{-}}\\
-\sqrt{\mathbb{C}}\sin r_2\,X_2+\cos r_2\,X_{\hat{-}}
\end{pmatrix}.
\end{align}
\begin{align}
x_{\hat{+}}'&=x_{\hat{+}}
 +\frac{\mathbb{S}\sin r_2}{\sqrt{\mathbb{C}}}x_2
 +\frac{\mathbb{S}(1-\cos r_2)}{\mathbb{C}}x_{\hat{-}}, \nonumber\\
x_{\hat{-}}'&=-\sqrt{\mathbb{C}}\sin r_2\,x_2
 +\cos r_2\,x_{\hat{-}}, \nonumber\\
x^{\hat{+}\prime}
&=\mathbb{C}x_{\hat{+}}'+\mathbb{S}x_{\hat{-}}'
=x^{\hat{+}}.
\end{align}

\reftitle{References}
\bibliography{BibTeXList}

\begin{thebibliography}{999}

\bibitem[Wess(1960)]{Wess1960}
Wess, J.
\newblock The conformal invariance in quantum field theory.
\newblock {\em Il Nuovo Cimento (1955-1965)} {\bf 1960}, {\em 18},~1086--1107.
\newblock {\url{https://doi.org/10.1007/BF02733168}}.

\bibitem[Kastrup(1966)]{Kastrup1966}
Kastrup, H.A.
\newblock Conformal Group in Space-Time.
\newblock {\em Phys. Rev.} {\bf 1966}, {\em 142},~1060--1071.
\newblock {\url{https://doi.org/10.1103/PhysRev.142.1060}}.

\bibitem[Mack and Salam(1969)]{SalamMack1969}
Mack, G.; Salam, A.
\newblock {Finite component field representations of the conformal group}.
\newblock {\em Annals Phys.} {\bf 1969}, {\em 53},~174--202.
\newblock {\url{https://doi.org/10.1016/0003-4916(69)90278-4}}.

\bibitem[Gross and Wess(1970)]{Gross1970}
Gross, D.J.; Wess, J.
\newblock Scale Invariance, Conformal Invariance, and the High-Energy Behavior
  of Scattering Amplitudes.
\newblock {\em Phys. Rev. D} {\bf 1970}, {\em 2},~753--764.
\newblock {\url{https://doi.org/10.1103/PhysRevD.2.753}}.

\bibitem[Weinberg(2010)]{Weinberg_2010}
Weinberg, S.
\newblock Six-dimensional methods for four-dimensional conformal field
  theories.
\newblock {\em Physical Review D} {\bf 2010}, {\em 82}.
\newblock {\url{https://doi.org/10.1103/physrevd.82.045031}}.

\bibitem[Brodsky et~al.(2015)Brodsky, de~Téramond, Dosch, and
  Erlich]{Brodsky_2015}
Brodsky, S.J.; de~Téramond, G.F.; Dosch, H.G.; Erlich, J.
\newblock Light-front holographic QCD and emerging confinement.
\newblock {\em Physics Reports} {\bf 2015}, {\em 584},~1–105.
\newblock {\url{https://doi.org/10.1016/j.physrep.2015.05.001}}.

\bibitem[Maldacena et~al.(2016)Maldacena, Stanford, and Yang]{Maldacena2016}
Maldacena, J.; Stanford, D.; Yang, Z.
\newblock Conformal symmetry and its breaking in two-dimensional nearly anti-de
  Sitter space.
\newblock {\em Progress of Theoretical and Experimental Physics} {\bf 2016},
  {\em 2016},~12C104.
\newblock {\url{https://doi.org/10.1093/ptep/ptw124}}.

\bibitem[Cunningham(1910)]{Cunningham1910}
Cunningham, E.
\newblock The Principle of Relativity in Electrodynamics and an Extension
  Thereof.
\newblock {\em Proceedings of the London Mathematical Society} {\bf 1910}, {\em
  s2-8},~77--98.
\newblock {\url{https://doi.org/https://doi.org/10.1112/plms/s2-8.1.77}}.

\bibitem[Bateman(1910)]{Bateman1910}
Bateman, H.
\newblock The Transformation of the Electrodynamical Equations.
\newblock {\em Proceedings of the London Mathematical Society} {\bf 1910}, {\em
  s2-8},~223--264.
\newblock {\url{https://doi.org/https://doi.org/10.1112/plms/s2-8.1.223}}.

\bibitem[Dirac(1936)]{Dirac1936}
Dirac, P.A.M.
\newblock Wave Equations in Conformal Space.
\newblock {\em Annals of Mathematics} {\bf 1936}, {\em 37},~429--442.

\bibitem[G{\"u}rsey(1956)]{Gürsey1956}
G{\"u}rsey, F.
\newblock On a conform-invariant spinor wave equation.
\newblock {\em Il Nuovo Cimento (1955-1965)} {\bf 1956}, {\em 3},~988--1006.
\newblock {\url{https://doi.org/10.1007/BF02823498}}.

\bibitem[Fulton et~al.(1962)Fulton, Rohrlich, and Witten]{Fulton1962}
Fulton, T.; Rohrlich, F.; Witten, L.
\newblock Conformal Invariance in Physics.
\newblock {\em Rev. Mod. Phys.} {\bf 1962}, {\em 34},~442--457.
\newblock {\url{https://doi.org/10.1103/RevModPhys.34.442}}.

\bibitem[Wilson(1969)]{Wilson1969}
Wilson, K.G.
\newblock Non-Lagrangian Models of Current Algebra.
\newblock {\em Phys. Rev.} {\bf 1969}, {\em 179},~1499--1512.
\newblock {\url{https://doi.org/10.1103/PhysRev.179.1499}}.

\bibitem[Kastrup(2008)]{Kastrup2008}
Kastrup, H.
\newblock On the advancements of conformal transformations and their associated
  symmetries in geometry and theoretical physics.
\newblock {\em Annalen der Physik} {\bf 2008}, {\em 520},~631--690.
\newblock {\url{https://doi.org/https://doi.org/10.1002/andp.200852009-1005}}.

\bibitem[Jackiw()]{Jackiw1990}
Jackiw, R., TWO-DIMENSIONAL CONFORMAL TRANSFORMATIONS REPRESENTED BY QUANTUM
  FIELDS IN MINKOWSKI SPACE-TIME.
\newblock In {\em Physics and Mathematics of Strings}; pp. 317--355.
\newblock {\url{https://doi.org/10.1142/9789814434461_0008}}.

\bibitem[Ji and Ravikumar(2026)]{Ji-Hari-2026-PRD}
Ji, C.R.; Ravikumar, H.
\newblock Interpolating conformal algebra in ($1+1$) dimensions between the
  instant form and the light-front form of relativistic dynamics.
\newblock {\em Phys. Rev. D} {\bf 2026}, {\em 113},~096018.
\newblock {\url{https://doi.org/10.1103/prlq-4j1l}}.

\bibitem[Di~Francesco et~al.(1997)Di~Francesco, Mathieu, and
  S{\'e}n{\'e}chal]{Francesco}
Di~Francesco, P.; Mathieu, P.; S{\'e}n{\'e}chal, D., Global Conformal
  Invariance.
\newblock In {\em Conformal Field Theory}; Springer New York: New York, NY,
  1997; pp. 95--110.
\newblock {\url{https://doi.org/10.1007/978-1-4612-2256-9_4}}.

\bibitem[Blumenhagen and Plauschinn(2009)]{Blumenhagen}
Blumenhagen, R.; Plauschinn, E., Basics in Conformal Field Theory.
\newblock In {\em Introduction to Conformal Field Theory: With Applications to
  String Theory}; Springer Berlin Heidelberg: Berlin, Heidelberg,  2009; pp.
  5--86.
\newblock {\url{https://doi.org/10.1007/978-3-642-00450-6_2}}.

\bibitem[Dirac(1949)]{Dirac1949}
Dirac, P.A.M.
\newblock Forms of Relativistic Dynamics.
\newblock {\em Rev. Mod. Phys.} {\bf 1949}, {\em 21},~392--399.
\newblock {\url{https://doi.org/10.1103/RevModPhys.21.392}}.

\bibitem[Ji and Mitchell(2001)]{Ji2001}
Ji, C.R.; Mitchell, C.
\newblock Poincar\'e invariant algebra from instant to light-front
  quantization.
\newblock {\em Phys. Rev. D} {\bf 2001}, {\em 64},~085013.
\newblock {\url{https://doi.org/10.1103/PhysRevD.64.085013}}.

\bibitem[Ji and Suzuki(2013)]{Ji2012}
Ji, C.R.; Suzuki, A.T.
\newblock Interpolating scattering amplitudes between the instant form and the
  front form of relativistic dynamics.
\newblock {\em Phys. Rev. D} {\bf 2013}, {\em 87},~065015.
\newblock {\url{https://doi.org/10.1103/PhysRevD.87.065015}}.

\bibitem[Ravikumar(2021)]{Ravikumar:2021xck}
Ravikumar, H.
\newblock {The Poincar{\'e} Algebra Interpolation between Instant Form Dynamics
  (IFD) and Light-Front Dynamics (LFD)}.
\newblock Master's thesis, National Institute of Technology Jalandhar,  2021.

\bibitem[Ji(2023)]{ji2023relativistic}
Ji, C.
\newblock {\em Relativistic Quantum Invariance}; Lecture Notes in Physics,
  Springer Nature Singapore,  2023.

\bibitem[Hornbostel(1992)]{Hornbostel1992}
Hornbostel, K.
\newblock Nontrivial vacua from equal time to the light cone.
\newblock {\em Phys. Rev. D} {\bf 1992}, {\em 45},~3781--3801.
\newblock {\url{https://doi.org/10.1103/PhysRevD.45.3781}}.

\bibitem[Ji and Rey(1996)]{Ji1996}
Ji, C.R.; Rey, S.J.
\newblock Light-front view of the axial anomaly.
\newblock {\em Phys. Rev. D} {\bf 1996}, {\em 53},~5815--5820.
\newblock {\url{https://doi.org/10.1103/PhysRevD.53.5815}}.

\bibitem[Ji et~al.(2015)Ji, Li, and Suzuki]{Ji2015EM}
Ji, C.R.; Li, Z.; Suzuki, A.T.
\newblock Electromagnetic gauge field interpolation between the instant form
  and the front form of the Hamiltonian dynamics.
\newblock {\em Phys. Rev. D} {\bf 2015}, {\em 91},~065020.
\newblock {\url{https://doi.org/10.1103/PhysRevD.91.065020}}.

\bibitem[Li et~al.(2015)Li, An, and Ji]{Ji2015SP}
Li, Z.; An, M.; Ji, C.R.
\newblock Interpolating helicity spinors between the instant form and the
  light-front form.
\newblock {\em Phys. Rev. D} {\bf 2015}, {\em 92},~105014.
\newblock {\url{https://doi.org/10.1103/PhysRevD.92.105014}}.

\bibitem[Ji et~al.(2018)Ji, Li, Ma, and Suzuki]{Ji2018QED}
Ji, C.R.; Li, Z.; Ma, B.; Suzuki, A.T.
\newblock Interpolating quantum electrodynamics between instant and front
  forms.
\newblock {\em Phys. Rev. D} {\bf 2018}, {\em 98},~036017.
\newblock {\url{https://doi.org/10.1103/PhysRevD.98.036017}}.

\bibitem[Ma and Ji(2021)]{Ji2021QCD}
Ma, B.; Ji, C.R.
\newblock Interpolating 't Hooft model between instant and front forms.
\newblock {\em Phys. Rev. D} {\bf 2021}, {\em 104},~036004.
\newblock {\url{https://doi.org/10.1103/PhysRevD.104.036004}}.

\bibitem[Ji(2025)]{Ji2025}
Ji, C.R.
\newblock Relativistic quantum invariance of {QED} and {QCD}.
\newblock {\em The European Physical Journal Special Topics} {\bf 2025}.
\newblock {\url{https://doi.org/10.1140/epjs/s11734-025-01803-9}}.

\bibitem[{'t Hooft}(1974)]{THOOFT1974461}
{'t Hooft}, G.
\newblock A two-dimensional model for mesons.
\newblock {\em Nuclear Physics B} {\bf 1974}, {\em 75},~461--470.
\newblock {\url{https://doi.org/https://doi.org/10.1016/0550-3213(74)90088-1}}.

\bibitem[Hepner(1962)]{Hepner1962}
Hepner, W.A.
\newblock The inhomogeneous Lorentz group and the conformal group.
\newblock {\em Il Nuovo Cimento (1955-1965)} {\bf 1962}, {\em 26},~351--368.
\newblock {\url{https://doi.org/10.1007/BF02787046}}.

\end{thebibliography}

\end{document}